%% file: main_paper.tex
\documentclass{article}

\usepackage[utf8]{inputenc} 
\usepackage[T1]{fontenc}    

\usepackage{amsmath,amsthm,alltt} 
\usepackage{graphicx,ctable,booktabs}
\usepackage{mathtools}
\usepackage{mathdots}
\usepackage{lipsum}

\usepackage{url}            
\usepackage{booktabs}       
\usepackage{amsfonts}       
\usepackage{nicefrac}       
\usepackage{microtype}      
\usepackage{xcolor}         
\usepackage{color-edits}
\addauthor{os}{blue}
\addauthor{bk}{red}
\addauthor{jm}{purple}

\usepackage[ruled,linesnumbered]{algorithm2e} 

\usepackage[margin=1in]{geometry}
\usepackage[numbers]{natbib}
\usepackage{float}
\usepackage{bbm}
\usepackage{diffcoeff}  
\usepackage{graphicx}
\usepackage{bm}
\usepackage{amsmath}
\usepackage{amssymb}
\usepackage{mathtools}
\usepackage{amsthm}

\usepackage{nccmath}

\usepackage{hyperref, cleveref}       
\input{macros}

\begin{document}
\title{Optimal Spend Rate Estimation and Pacing\\for Ad Campaigns with Budgets}
\author{Bhuvesh Kumar\and Jamie Morgenstern \and Okke Schrijvers}
\date{}
\maketitle

\begin{abstract}
    Online ad platforms offer budget management tools for advertisers that aim to maximize the number of conversions given a budget constraint. As the volume of impressions, conversion rates and prices vary over time, these budget management systems learn a spend plan (to find the optimal distribution of budget over time) and run a pacing algorithm which follows the spend plan.
    
    This paper considers two models for impressions and competition that varies with time: a) an episodic model which exhibits stationarity in each episode, but each episode can be arbitrarily different from the next, and b) a model where the distributions of prices and values change slowly over time. We present the first learning theoretic guarantees on both the accuracy of spend plans and the resulting end-to-end budget management system. We present four main results: 1) for the episodic setting we give sample complexity bounds for the spend rate prediction problem: given $n$ samples from each episode, with high probability we have $\abs{\widehat{\rho}_e - \rho_e} \leq  \widetilde{O}\br{\frac{1}{n^{1/3}}}$ where $\rho_e$ is the optimal spend rate for the episode, $\widehat{\rho}_e$ is the estimate from our algorithm,
    2) we extend the algorithm of \citet{balseiro2019learning} to operate on varying, approximate spend rates and show that the resulting combined system of optimal spend rate estimation and online pacing algorithm for episodic settings has regret that vanishes in number of historic samples $n$ and the number of rounds $T$, 3) for non-episodic but slowly-changing distributions we show that the same approach approximates the optimal bidding strategy up to a factor dependent on the rate-of-change of the distributions and 4) we provide experiments showing that our algorithm outperforms both static spend plans and non-pacing across a wide variety of settings.
\end{abstract}

\section{Introduction}
\label{sec:intro}

\input{intro}

\section{Setting and Preliminaries}
\label{sec:setting}
\input{setting}

\section{Approximating Optimal Spend Rates}
\label{sec:spend_prediction}
\input{spend_pred}

\section{Pacing using Approximate Spend Rates}
\label{sec:pacing}

\input{online_pacing}

\section{Slow-moving Distributions}
\label{sec:slow_changing}
\input{slowly_changing}

\section{Experiments}
\label{sec:experiments}
\input{experiments}

\bibliographystyle{plainnat}
\bibliography{references}

\clearpage
\newpage
\appendix

\input{appendix}

\end{document}

%% file: macros.tex
\newcommand{\bc}[1]{\left\{{#1}\right\}}
\newcommand{\br}[1]{\left({#1}\right)}
\newcommand{\bs}[1]{\left[{#1}\right]}
\newcommand{\abs}[1]{\left| {#1} \right|}
\newcommand{\norm}[1]{\left\| {#1} \right\|_{\infty}}
\newcommand{\ceil}[1]{\left\lceil #1 \right\rceil}
\newcommand{\floor}[1]{\left\lfloor #1 \right\rfloor}

\DeclarePairedDelimiter{\defaultDelim}{[}{]}

\newcommand{\E}[2][]{\mathbf{E}_{#1}\defaultDelim*{#2}}

\newcommand{\proj}[2][]{\textsc{Proj}_{[#1]}\defaultDelim*{#2}}

\newcommand{\ind}[1]{\mathbbm{1}\bc{#1}}
\newcommand{\pp}[2]{\pi^{#1}\br{#2}}
\newcommand{\ppe}[2]{\pi_e^{#1}\br{#2}}

\newcommand{\pos}[1]{\br{#1}^+}

\newcommand{\ereg}[2]{\epsilon\textsc{-Reg}(#1; #2)}
\newcommand{\areg}[2]{\alpha\textsc{-Reg}(#1; #2)}
\newcommand{\omereg}[2]{(1-\epsilon)\textsc{-Reg}(#1; #2)}

\def\mut{\mu^\star}
\def\mute{\mu_e^\star}

\def\osr{\rho}

\def\hind{H}

\def\te{\tau}

\def \vmax{h}

\newcommand{\duale}[1]{\Psi_{e}(#1)}

\newcommand{\dualegrad}[2]{\diffp{\Psi_{e}(\mu, #2)}{\mu}[#1]}
\newcommand{\spende}[1]{\overline{G_e}(#1)}

\newcommand{\spend}[1]{\overline{G}(#1)}
\newcommand{\espend}[1]{\widehat{G}(#1)}
\newcommand{\espende}[1]{\widehat{G_e}(#1)}
\newcommand{\spendef}[0]{\overline{G_e}}
\newcommand{\spendf}[0]{\overline{G}}
\newcommand{\espendf}[0]{\widehat{G}}
\newcommand{\espendef}[0]{\widehat{G_e}}

\newcommand{\approxspend}{{\small \textsf{ApproxSpendSP}}\xspace}
\newcommand{\approxspendfp}{{\small \textsf{ApproxSpendFP}}\xspace}
\newcommand{\approxspendrate}{{\small \textsf{ApproxSpendRate}}\xspace}
\newcommand{\adaptivepacing}{{\small \textsf{EpisodicAdaptivePacing}}\xspace}

\def\v{\bm{v}}
\def\p{\bm{p}}

\newtheorem{theorem}{Theorem}
\newtheorem{definition}[theorem]{Definition}
\newtheorem{example}[theorem]{Example}
\newtheorem{lemma}[theorem]{Lemma}

\theoremstyle{definition} 
{}

%% file: intro.tex
Online advertising is a massive industry worth around $\$140$ billion dollars in 2020 in the United States alone~\citep{statista2020}. Advertisers bidding within large online platforms are usually constrained by budget, and must decide how to distribute this budget over time as the supply and demand of impressions change. For example, there are more users online during the day than at night, leading to a variable density of impressions opportunities (see e.g. Figure 1 in \citet{liu2020moment}). Furthermore, users may be more likely to interact with an ad outside of working hours, leading to those impressions generating more value for advertisers (e.g. Table 2 in \citet{liu2020moment}). Finally, competition for impressions  may vary over the course of the day, as  other advertisers may allocate more budget to high-value periods (e.g. Figures 2 and 3 in \citet{agarwal2014budget}).

The temporal effects have led to a variety of work on constructing \emph{spend plans} for a campaign which learn how to distribute a budget over time~\citep{ma2019large, agarwal2010forecasting, agarwal2014budget, lee2013real}. Generally, the approach taken in these works is twofold: first, they use some model (e.g. a high dimensional time series model) to forecast the number of impression opportunities over the course of a day. This is taken as the spend plan. Secondly, they use a \emph{pacing algorithm}, which tries to match the empirical spend rate to the spend plan.  \citet{lee2013real} modify the bid to control spend, while \citet{agarwal2014budget} modify their participation probability   to control spend. 

There are several limitations to the above approaches. First, they model the density of impression opportunities  assuming that value per user and price per user is roughly constant. Considerable evidence ~\citep{liu2020moment,agarwal2014budget} refutes this assumption, suggesting that conversion rates and prices change over time. Second, their work focuses on empirical rather than theoretical results, limiting our understanding about which settings we can predict the resulting algorithms to have good performance. This motivates the problem that we study in this paper: \emph{Can we identify non-stationary settings for which we can provably learn a spend plan that approximates the optimal distribution of budget, and where the end-to-end system provably performs well?}

More formally: We study the problem of computing optimal spend plans from a learning-theoretic perspective in two settings: an episodic model, and a model in which price and value distributions change smoothly over time. For the first, we consider an advertiser with budget $B$ that participates in a sequence of $T$ single-item second-price auctions, called rounds. These auctions are divided into $E$ episodes of $\tau = \tfrac{T}{E}$ rounds\footnote{We assume equal sized episodes to simplify the presentation of the paper. Our results can be generalized to different sized episodes where the size of episodes can also be estimated.}. Each episode $e \in [E]$ has a fixed product distribution $Q_e = F_e \times D_e$, with values $v_t \sim F_e$ for $v_t \in \bs{0,\vmax}$ and independently prices $p_t \sim D_e$  in $p_t \in \mathbbm{R}^+$. Prices and values within an episode are i.i.d., while prices and values across episodes are independently, but not identically, distributed.
Let $\rho = \frac{B}{T}$ be the \emph{average spend per round} of a strategy spending budget $B$ over $T$ rounds. For all $e$, $f_e$ and $d_e$ denote the probability density functions (pdf) of distributions $F_e$ and $D_e$ respectively. 
Second, we consider a non-episodic setting, where all distributions are guaranteed to change smoothly: each round has a product distribution $Q_t = F_t \times D_t$ with the property that $\norm{F_{t+1} - F_{t}} \leq \zeta$, and $\norm{d_{t+1} - d_t} \leq \theta$  for all $t\in [T]$.

For both settings, we ask: First, can we accurately estimate an optimal spend allocation?  Second, given an (approximately) optimal spend plan, can we implement a pacing algorithm that satisfies the budget constraint and achieves vanishing regret compared to the ex-post optimal?

\subsection{Main Contributions}
Our main contributions are as follows.
\begin{itemize}
    \item {\bf Episodic Setting}. We propose a pair of algorithms 1) \approxspendrate, an offline algorithm that estimates the optimal spend plan on $n$ samples, and 2)
        \adaptivepacing, an online algorithm that adaptively follows the spend plan over $T$ new auctions,
        that jointly have regret vanishing in $n$ and $T$, compared to the best bidding strategy in hindsight.
        The formal statement appears as Theorem~\ref{thm:main_theorem_formal} in Section~\ref{sec:pacing} and relies on the following additional results:
        \begin{itemize}
            \item {\bf Estimating Optimal Spend Plan.}
            In Section~\ref{sec:spend_prediction} we bound the accuracy of constructing of an optimal spend plan. We give an algorithm \approxspendrate, that given $n$ samples from each episode, with probability at least $1-\frac{2E}{n}$, produces a spend plan that satisfies $\abs{\widehat{\rho}_e - \rho_e} \leq (E+1)  \cdot \widetilde{O}\br{\frac{1}{n^{1/3}}}$ where $\rho_e$ is the optimal spend rate for the episode, $\widehat{\rho}_e$ is the estimate from our algorithm and $E$ is the number of episodes.
            \item {\bf Online Pacing Algorithm on Spend Plan.}
            In Section~\ref{sec:pacing} we then give an adaptive pacing algorithm \adaptivepacing that takes an (approximately accurate) spend plan, and implements a bidding strategy that follows this spend plan over $T$ new auctions. The regret of this algorithm vanishes in $n$ and $T$ with respect to the best bidding strategy in hindsight.
        \end{itemize}
    \item {\bf Slow-moving Distributions}. In Section~\ref{sec:slow_changing}, for slow-moving distributions we learn a spend plan as if the data came from an episodic model with number of episodes $E$. The end-to-end performance achieves a constant factor approximation of the to the best bidding strategy in hindsight, where the constant factor depends on the rate at which the distributions change.
    \item {\bf Experimental Evidence}. Finally, in Section~\ref{sec:experiments} we present experiments where we compare the performance of our method to the ~\citet{Balseiro:2017} algorithm (which neither estimates nor uses a spend plan as it was designed for adversarial and stationary settings). Our method compares favorably to the ex-post optimal strategy and outperforms other methods in a wide variety of settings. 
\end{itemize}

\subsection{Related Work}
\paragraph{Optimal Spend Rate Estimation.} 
There are number of works that aim to estimate optimal spend rates for budget pacing \citep{ma2019large, agarwal2010forecasting, agarwal2014budget, lee2013real}. \citet{ma2019large} and \citet{agarwal2010forecasting} primarily focus on the on the spend plan estimation. Both of these papers aim to forecast user visits, which correlates strongly with the number of impression opportunities. They do this using time series modeling techniques for users within the targeting criteria of a campaign. These works do not attempt to estimate how conversion rates or prices for ad opportunities change over time. \citet{lee2013real} and \citet{agarwal2014budget} combine user visit estimates with an online pacing algorithm to match the spend rate to the user visit rate. Similar to the approaches below, \citet{lee2013real} uses a multiplicative shading strategy (i.e. bidding $\alpha \cdot v_t$ instead of $v_t$) to control spend, while \citet{agarwal2014budget} participate in each auction with a parameterized probability to control spend. None of these papers give formal guarantees on the performance of the end-to-end budget management system.

{\bf Online Algorithms for Pacing.} Work on pacing algorithms has only focused on guarantees for pacing algorithms in absence of a spend plan. In many cases, for repeated second-price auctions, the optimal pacing strategy in hindsight is a multiplicative shading strategy (i.e. bidding $\alpha \cdot v_t$ for the auction at time $t$ for a fixed $\alpha \le 1$ that does not vary over time) \citep{rusmevichientong2006, feldman2007budget, hosanagar2008, balseiro2019learning, conitzer2018pacing}. \citet{balseiro2019learning} were the first to give online learning algorithms that approximate this best response. For i.i.d. value and price distributions, they give an online algorithm with  regret $O(T^{1/2})$. Similar guarantees are also shown by \citet{balseiro2020dual} who achieve $O(T^{1/2})$ regret for  stationary value and price distributions setting without assuming independence between values and prices. 
There are few works that give provable guarantees for non-stationary competition and values. \citet{balseiro2019learning} consider the case of adversarial values and prices and show that no algorithm can achieve sub-linear regret with respect to any benchmark that obtains more than $\frac{B}{T h}$ fraction of the utility obtained by using the optimal strategy with the power of hindsight (where $h$ is an upper bound on the value). They also give an algorithm which obtains the $\tilde{O}(T^{1/2})$ upper bound on the regret with respect to  $\frac{B}{T h}$ fraction of the optimal. \citet{balseiro2020dual} considers both an ergodic setting and a periodic setting where regret grows as $\tilde{O}(T^{1/2})$. Their algorithms do not construct a spend plan and instead rely on the fact that at a macro-level the expected optimal spend rate is constant. By contrast, in our setting obtaining no-regret may depend on saving enough budget for the end of the campaign (for example to reach users on the weekend for a week-long campaign). Only by explicitly constructing an approximately optimal spend plan can one give guarantees for such campaigns.

\citet{ConitzerKPSSSW2019} show that for individual first-price single-item auctions, multiplicative shading yields the Eisenberg-Gale outcome of the corresponding Fisher market (though generally multiplicative shading is not a best response in this setting). \citet{gao2021online} give an online learning algorithm for this setting that results in this equilibrium and can be run in a decentralized way by each advertiser individually.

While bid modification yields the an optimal strategy for a bidder, an alternative way to respect a budget constraint is to limit the number of auctions a bidder participates in. \citet{mehta2007adwords} give revenue guarantees for the online matching problem where users (in this case, impressions for sale) arrive one at a time and the auction selects a winner who pays their bid; once a bidder has exhausted their budget they will no longer be selected as a winner. Subsequently bidder selection has been applied to more general settings \citep{abrams2008comprehensive,azar2009gsp,goel2010gsp,karande2013pacing}. Since truthful bidding is not a best response for advertisers in bidder selection mechanisms, this line of work is less directly relevant to our work.

In the previous two lines of work, advertisers know the value they have for an impression when they bid. A separate line of work considers a bandit setting, where the value is only revealed to advertisers after they win an auction.~\citet{amin2012budget} and~\citet{tran2012knapsack} give theoretical guarantees for discrete value distributions. \citet{flajolet2017real} extend these results to continuous distributions. Finally, \citet{aless2020online} and \citet{avadhanula2020} consider the problem of allocating budget across different channels. The different channels have different distributions and as such bear some similarity to the setting we consider. However, since all channels are simultaneous available and each channel is i.i.d., the spend rate remains constant over time (cf. our setting where spend rates change).

\paragraph{Equilibrium Analysis}
In addition to the work on online algorithms, there's a growing body of work that analyzes the equilibria of pacing systems under the assumption that all advertisers use the same bid-shading approach, e.g. \cite{cray2007,borgs2007dynamics,balseiro2017budget, conitzer2018pacing, ConitzerKPSSSW2019, babaioff2020non, chen2021complexity}. The framework of \citet{balseiro2020budget} studies stationary equilibria and characterize Bayesian optimal mechanisms that satisfy budget constraints.


%% file: setting.tex
We study the problem of designing a bidding algorithm for budget-constrained advertisers in non-stationary settings. This bidding algorithm aims to maximize utility subject to a given budget constraint $B$.  The algorithm participates in a sequence of $T$ single-item second-price auctions\footnote{Our model  captures additional settings, including posted prices and second-price auctions with reserve prices, but for ease of exposition we consider second-price auctions throughout.}. We refer to each auction as a round. In the following we present the notation for the episodic setting that we study, the non-episodic setting is formally introduced in Section~\ref{sec:slow_changing}.

In every round $t \in[T] $, the bidder observes a value $v_t$ for the impression opportunity\footnote{ $v_t$ could capture the value  $v$ that the advertiser has for a conversion times the probability of a conversion of the impression opportunity. The latter may depend on context like the user or a search query and is estimated by the platform.} and submits a bid $b_t$ to the auctioneer.  Let $p_t$ be the highest competing bid for the impression opportunity. When $b_t \ge p_t$ the bidder wins, spends $p_t$, and gains utility $u_t = v_t - p_t$. Otherwise she loses, pays $0$, and gains utility $u_t = 0$. The bidder's goal is to maximize their utility subject to the sum of expenditures across $T$ rounds being at most $B$. 

A strategy $\sigma$ of the bidder is a sequence of deterministic\footnote{While all of our algorithms are deterministic, the lower bound in Lemma~\ref{lem:hist_lb} can be extended to randomized mappings as well. For ease of exposition, our algorithms
use only deterministic strategies.} mappings $\sigma_1, \ldots, \sigma_T$ where $\sigma_t$ uses the information that is available to bidder in round $t$ to produce bid $b^\sigma_t$. We focus attention on strategies that respect the budget constraint.

\begin{definition}[Budget-feasibly Strategy]
\label{def:budget_feasible}
$\sigma$ is budget feasible if $\sum_{t=1}^T \ind{b^\sigma_t \geq p_t}p_t \le B$
 for any realized values $\v = v_1, \ldots , v_T$ and prices $\p = p_1, \ldots, p_T$.
\end{definition}

A strategy's utility is simply its total utility over $T$ rounds.

\begin{definition}[Performance of a Strategy]
\label{def:performance}
For a given budget feasible strategy $\sigma$, it's performance on a realized sequence of values $\v$ and prices $\p$ is given by 
\begin{equation}
    \pp{\sigma}{\v ; \p} = \sum_{t=1}^T \ind{b^\sigma_t > p_t}(v_t - p_t).
\end{equation}
\end{definition}

As benchmark, we consider the best (fractional) allocation in hindsight on the realized values $\v$ and prices $\p$. While the benchmark may appear to be strong, it is commonly used in the budget pacing literature.

\begin{definition}[Hindsight Strategy Benchmark]
\label{def:benchmark}
The performance of the hindsight strategy $H$ on a realized sequence of values $\v$ and prices $\p$ is given by 
\begin{align*}
    \pp{\hind}{{\v ; \p}} = \max_{x \in [0,1]^T} & \sum_{t=1}^T (v_t - p_t)x_t  \hspace{.1in}  \text{s.t.} \hspace{.1in} \sum_{t=1}^T p_tx_t \leq B 
\end{align*}
\end{definition}
Here $x_t \in [0,1]$ represents a fractional allocation of impression opportunities.
 We measure the regret of a strategy compared to the benchmark in expectation over the values and prices. We use the notion of $\alpha$-regret proposed by \citep{kakade2009playing}, a multiplicative notion:
\begin{definition}[$\alpha$-regret]
For strategy $\sigma$ and $\alpha \in (0,1]$, the $\alpha$-regret with respect to the hindsight strategy is:
\begin{equation*}
    \areg{\sigma}{T} = \alpha{\E[]{\pp{\hind}{\v,\p}}} -{\E[]{\pp{\sigma}{\v,\p}}}
\end{equation*}

\end{definition}
Where the expectation is over $(\v,\p)$ sampled from $\vec{Q}$, that is,  $(v_t, p_t)$ in episode $e$ is sampled from $Q_{e} = F_e \times D_e$.Our algorithm first constructs a \emph{spend plan} prior to the $T$ auctions, using historical data. The accuracy of the spend plan will be a function of the sample size our algorithm is given.

\begin{definition}[Sample Complexity]
The sample complexity of achieving a given approximation factor $1-\epsilon$ is the minimum number of samples $m$ such that there exists an (offline) learning algorithm $A$ with the desired approximation.
\end{definition}
Of particular interest are algorithms where both the $\alpha$-regret is sublinear in $T$, and additionally, $\alpha$ approaches $1$ using a polynomial number of samples. We overload the term ``vanishing regret'' for such situations.
\begin{definition}[Vanishing Regret]
A strategy $\sigma_n$ (which has access to $n$ samples from $\vec{Q}$) has \emph{vanishing regret} if $\omereg{\sigma_m}{T} = o(T)$ \emph{and} $m \in O(\text{poly}(\epsilon^{-1}))$.
\end{definition}

\subsection{Outline of the Solution}
As mentioned previously, our algorithm first produces a spend plan from data, then uses a pacing algorithm to meet that spend plan. The former is an offline learning problem that happens before the campaign starts. The latter is an online algorithm that operates on the spend plan and realized expenditures. Before going into these components, it is informative to understand why this decomposition in a spend plan and pacing algorithm makes sense.

\paragraph{Why Historical Data is Needed.}
 \citet{balseiro2020budget} have studied pacing for non-stationary distributions without using historical data. Could the episodic setting that we're studying be amenable to positive results without historical data too? Unfortunately this is not the case. The following is an example with two episodes for deterministic algorithms. We generalize the example in the lemma that follows.

\begin{example}\label{ex:data}
Consider two instances of the episodic setting characterized by the episodic distributions, $I = (Q_1,Q_2)$ and $I' = (Q_1,Q'_2)$ (where $Q=F\times D$). All the distributions consist of a single atom: prices distributions $D_1 = D_2 = D'_2$ and yield $1$ with probability $1$. The value generated by $F_1$ is $2$, by $Q_2$ is $1$ and by $Q'_2$ is $3$.

Consider a buyer with budget $B=\tfrac12T$, thus they can buy precisely half the impression opportunities. In both instances, the first episode yields utility $2-1 = 1$ per round that is won, but the second episode differs for the two instances. For $I$, the per-round utility when the bidder wins is $1-1 = 0$, while for $I'$ it is $3-1 = 2$.

So if the bidder faces the first instance, she needs to win all but a sublinear (in $T$) number of rounds in episode 1 for vanishing regret, but if she faces the second instance she may win at most a sublinear number of rounds in the first episode. Since she doesn't know which instance she faces until she enters episode 2, any strategy must incur $\Omega(T)$ regret on at least one of $I$, $I'$.
\end{example}

The example above can be generalized to a stronger result for instances with more episodes and that includes randomized algorithms.

\begin{lemma}
\label{lem:hist_lb}
Any strategy $\sigma$ that only depends on the history $\mathcal{H}_t = (v_i,b_i,p_i)_{i=0}^{t-1} \cup v^t$ in round $t$, for a large enough $T$, and budget $B$ such that $0 < \osr = \frac{B}{T} < \vmax$, for any number of episodes $E$, for any $\epsilon$ such that $1 - \epsilon > \max \{\frac{\rho}{\vmax}, \frac{1}{E}\}$, there exists an instance of the episodic setting with distributions  $\vec{Q} = (Q_1, \cdots, Q_E)$ such that
$$ \omereg{\sigma}{T} \geq \Omega(T).$$
\end{lemma}
\begin{proof}
In an adversarial setting where the values and prices are arbitrary, in each round $t$, \citet [Theorem 1] {balseiro2019learning} show that for any strategy $\sigma$ such that $\sigma_t$ depends only on the history $\mathcal{H}_t$, for a large enough $T$, for any budget $B$ satisfying $\osr = \frac{B}{T} < \vmax$ where $\vmax$ is the upper bound on the values, for any $\epsilon$ such that $(1 - \epsilon) > \frac{\osr}{\vmax}$, there exists adversarial values $\v$ and $\p$ such that $ (1 - \epsilon) \pp{\hind}{\v ; \p} - \pp{\sigma}{{\v ; \p}} \geq \Omega(T)$. 

Note that if $\osr \geq \vmax$ then truthful bidding is feasible and achieves the optimal utility. When $\osr < \vmax$, then the above result says that there exists a barrier of $\frac{\osr}{\vmax}$ such that for any strategy $\sigma$ that only depends on the history, there always exists an instance where $\sigma$ cannot obtain better than $\frac{\osr}{\vmax}$ fraction of the optimal utility. In other words, if the budget constraint is active,  the lower the budget, the smaller the fraction of the optimal budget constrained utility the advertiser can hope to attain.

Since we don't make any assumptions about how distributions $Q_e$ are related across the episodes, we can extend the analysis and the adversarial case example of \citet[Theorem 1]{balseiro2019learning} to work in the episodic setting and show that similar lower bounds can be shown for such strategies in our setting as well. 
Specifically, in the proof of \citet[Theorem 1]{balseiro2019learning}, the adversarial example has the value $v_t$ fixed as $\vmax$ for all $T$ rounds, and the price profile is samples from a distribution such that the $T$ round auction is divided into $m$ \emph{episodes}. When $\frac{\vmax}{E} \leq \osr = \frac{B}{T} < \vmax$, then by setting $m = E$ in the proof for \citet[Theorem 1]{balseiro2019learning}, we can recover the guarantee that there exists an instance such that $ \ereg{\sigma}{T} \geq \Omega(T)$ for $1 - \epsilon \geq \frac{\osr}{\vmax}$. If $0 < \rho < \frac{\vmax}{E}$, then the complete example is scaled down by replacing $\vmax$ with $\vmax' = \osr E$. Note that since $\rho < \frac{\vmax}{E}$, we have that $\vmax' < \vmax$, thus the example is valid. We also get that $\frac{\vmax'}{E} \leq \osr = \frac{B}{T} < \vmax'$. Thus, we obtain that $ \ereg{\sigma}{T} \geq \Omega(T)$ for $1 - \epsilon \geq \frac{\osr}{\vmax'} = \frac{1}{E}$.
\end{proof}

Since algorithms in the episodic setting that only operate on the immediate history fail to have vanishing regret, we use access to historical data in the form of samples from the distribution.

\paragraph{Why Spend Plans are Needed.}
With access to samples from the distribution, one could still attempt to design an algorithm that does not involve a spend plan. Recall from the related work that ex-post the optimal bidding strategy is to bid $\beta^*\cdot v_t$ for some constant $\beta^*$. So what if we used samples to estimate this $\beta$ and used this directly? The following lemma shows that this yields linear regret with constant probability.

\begin{lemma}
\label{lem:ex-ante}
There exists an instance of the episodic setting with distributions $\vec{Q} = (Q_1, Q_2)$ such that the ex-ante optimal pacing multiplier $\beta^*$ incurs $O(T)$ regret with probability at least $\tfrac19$.
\end{lemma}
\begin{proof}

Let $Q_1 = F_1 \times D_1$ and  $Q_2 = F_2 \times D_2$ such that the prices drawn from $D_1$ in episode 1 are equal to some positive constant $p_{\text{high}}$ and the prices drawn from $D_2$ in episode $2$ is equal to $p_{\text{low}}$ such $p_{\text{low}} = \frac{p_{\text{high}}}{\te}$. 
The values drawn from $F_1$ are $0$ with probability $\tfrac{\te - 1}{\te}$ and are equal to $p_{\text{high}} + v_{\text{low}}$ with the remaining probability $\tfrac{1}{\te}$ for some constant $v_{\text{low}}$. The values drawn from $F_2$ in the second episode is always equal to $p_{\text{low}} + v_{\text{high}}$ such that $ v_{\text{high}} > v_{\text{low}}$.

Consider the episodic setting defined by distribution $\vec{Q} = (Q_1, Q_2)$ and budget $B = 2p_{\text{high}}$. In episode $1$, most of the impressions have no value but with a small probability, we might observe high value and expensive impressions. In the episode $2$, every impression is cheap and high value.

The corresponding ex-ante optimal pacing multiplier $\beta^*$ in this case is $\frac{p_{\text{high}}}{p_{\text{high}} + v_{\text{low}}}$. The pacing strategy derived from $\beta^*$ always bids $b_t = \beta^* v_t$. In expectation, $\sigma^{\beta^*}$ wins all impressions in episode $2$ each with utility $v_{\text{high}}$, and wins $1$ impression of utility $v_{\text{high}}$ resulting in expected utility of $\frac{T}{2}v_{\text{high}} + v_{\text{low}}$.

Although the expected utility is high, but in the realized case, if more than one impression of value $p_{\text{high}} + v_{\text{low}}$ and price $p_{\text{high}}$ is present in episode $1$, then the fixed pacing strategy wins both these impressions with depletes the complete budget, leaving no budget for episode 2 and a total utility $\pp{\beta^*}{\v,\p} = 2v_{\text{low}}$. When compared to the Hindsight strategy $H$, the hindsight strategy always ends up buying all the impressions in second episode and thus gets utility  $\pp{\hind}{\v,\p} \geq \frac{T}{2}v_{\text{high}}$.

Thus, if at least two impressions with positive utility appear in episode $1$, the regret $\pp{\hind}{\v,\p} - \pp{\beta^*}{\v,\p}$ is equal to  $\frac{T}{2}v_{\text{high}} - 2v_{\text{low}}$ which is $O(T)$. The probability that at least two positive utility impressions appear in episode $1$ is $1 - 2(\frac{\te-1}{\te})^{\te-1}$ which is greater than $\frac{1}{9}$ for $\te \geq 3$.

This implies that with probability at least $\frac{1}{9}$, $\pp{\hind}{\v,\p} - \pp{\beta^*}{\v,\p} = O(T)$
\end{proof}

Instead we construct an intermediate spend plan and combine this with an adaptive pacing algorithm; an approach we outline next.
 
\paragraph{Outline of the Solution using Spend Plans.} 

To understand the optimal spend plan, we first introduce the notation of spend functions which represent the expected expenditure of strategies that shade by a fixed shading multiplier. 

\begin{definition}[Spend Function]
\label{def:spend_function} 
Consider a fixed pacing strategy $\sigma_\mu$ that always bids $b_t = \frac{1}{\mu +1}^\cdot v_t$ and is not restricted by any budget constraints. The expected expenditure of $\sigma_\mu$ in a single round in episode $e$ is
\begin{align}
\tiny 
\spende{\mu} &= \E[v \sim F_e,  p \sim D_e]{\ind{v \geq (1 + \mu) p} p} \\
&= \int_0^\vmax (1 - F_e((1 + \mu)p)) \cdot p \cdot d_e(p) dp.
\label{eq:spend_fn_def}
\end{align}
\end{definition}

\begin{definition}[Optimal Spend Rates]
Given an episodic setting where the ex-post optimal bidding strategy is to bid $\beta^*\cdot v_t$, we define $\bm{\rho} = \rho_1, \cdots, \rho_E$ as optimal spend rates if for all $e$, $\rho_e = \spende{\mut}$, where $\frac{1}{1 + \mut} = \beta^*$.
\label{def:optimal_spend_rates}
\end{definition}
In simpler words, the optimal spend plan is characterized by the optimal spend rates  $\bm{\rho} = \rho_1, \cdots, \rho_E$, such that the $\rho_e$ is equal to expected expenditure of the ex post optimal bidding strategy in a single round in episode $e$. We define the dual of the expectation of the optimization problem in Definition~\ref{def:benchmark} as
\begin{equation}
    \Psi(\mu) =  \E[\v,\p]{\sum_{t=1}^T\pos{v_t - (1 + \mu) p_t}  + \mu B  }.
\end{equation}
The ex-post optimal bidding strategy  $\sigma_{\mut}$ bids $\frac{v_t}{1 + \mut}$ where $\mut$ is the dual minimizer ($ \Psi(\mut)  = \inf_{\mu \geq 0} \Psi (\mu)$), which spends the complete budget in expectation:
\begin{equation}
   \tau \sum_{e=1}^E \spende{\mut} =  \tau \sum_{e=1}^E \rho_e = B,
   \label{eq:spend_all}
\end{equation}
where  $\rho_1, \cdots, \rho_E$ are the optimal spend rates. Refer to Appendix~\ref{app:optimal_b} for a detailed exposition about the characterization of the optimal spend rates through the dual of the problem. Knowing the optimal spend rates can help decompose the entire campaign into smaller budget constrained campaigns for each episode where the distributions of values and prices remain stationary. The exact formulation of optimal spend rates requires complete knowledge of the distributions $\vec{Q} = (Q_1, \cdots, Q_E)$ which is not available, instead we have access to historic samples from the distribution $\vec{Q}$. This is reasonable to assume as we are designing this framework for large online ad exchanges which usually have access to a lot of historical data. Our solution is is a two step pipeline:
\begin{enumerate}
    \item \textbf{1. Approximate optimal spend rates:} Use historical samples from $\vec{Q}$ to approximate optimal spend rates $\bm{\rho} = \rho_1, \cdots, \rho_E$ as $\bm{ \widehat{\rho}} =  \widehat{\rho}_1, \cdots, \widehat{\rho}_E$.
     \item \textbf{2. Adaptive pacing with spend rates:} Use the approximate spend rates to construct an online pacing algorithm that runs on realized impressions.
\end{enumerate}

\subsection{Preliminaries}

\label{app:concentration}
We will use some results on uniform convergence and pacing for i.i.d. settings in this paper.

\subsubsection{Dvoretzky-Kiefer-Wolfowitz (DKW) Inequality}
The Dvoretzky-Kiefer-Wolfowitz (DKW) inequality \citep{dvoretzky1956asymptotic,massart1990tight} gives a uniform convergence bound on the empirical cumulative distribution function.

\begin{lemma}[DKW Inequality]
\label{lemma:dkw}
Given $n$ samples $X_1, X_2, \ldots, X_n$ from a distribution $F$. The empirical cdf on the samples is given by $\widehat{F}(x) = \frac{1}{n} \sum_{i=1}^n \ind{X_i < x}$. With probability at least $1 - \delta$
$$\norm{F - \widehat{F}} \leq \sqrt{\tfrac{\log \tfrac{2}{\delta}}{2n}}.$$
\end{lemma}

\subsubsection{Kernel Density Estimation} While the DKW inequality~\ref{lemma:dkw} gives strong uniform convergence bounds on the cdf of a distribution, bounding the probability density function (pdf) of a distribution is more challenging. A common approach to do this is to use Kernel Density Estimation (KDE) \citep{davis2011remarks, parzen1962estimation}. Let $D$ be the distribution with pdf $d$ that we want to estimate as $\widehat{d}$. Formally the Kernel Density Estimation is defined below for scalar distributions which we use in our setting. 

\begin{definition}[Kernel Density Estimation] Given a kernel $K$, scalar $s$, and $n$ samples $X_1, \cdots, X_n$ from the distribution $D$, the KDE is given by 
$$\widehat{d}(x) = \frac{1}{n\cdot s} \sum_{i=1}^n K\left(\frac{x - X_i}{s}\right).$$
\end{definition}

While most results on KDE are for bounding the mean squared error ($\E[d]{(\widehat{d} - d)^2}$), recent work by \citet{jiang2017uniform} gives a uniform convergence guarantee for KDE. We present a simplified version of their result below.

\begin{lemma}[\citep{jiang2017uniform}]
\label{lem:jiang}
If $d$ is Lipschitz and bounded i.e. there exists a constant $C_1$ such that $|d(x) - d(x')| \leq C_1\abs{x- x'}$ for $x, x' \in \mathbbm{R}$ and $\norm{d} \leq C_2$ for some constant $C_2$, then there exists a constant $C'$ (that depends on $K$, $C_1$, $C_2$, and some other constants), such that with probability at least $1 - 1/n$ and by setting $s = n^{-1/3}$, the kernel density estimate $\widehat{d}$ satisfies that
\begin{align*}
    \norm{\widehat{d} - d} &\leq C'\br{s + \sqrt{\tfrac{\log n}{ns}}} = \widetilde{O}\br{\frac{1}{n^{1/3}}}.
\end{align*}
provided that $K$ is spherically symmetric, non-increasing, and has exponential decay (i.e. $K(x) = k(\abs{x})$ where $k:\mathbbm{R}^+ \rightarrow \mathbbm{R}^+$ is a non decreasing function s.t. for all $u > u_{\eta}$, $k(u) \leq C_{\eta} \exp({-u^{\eta}})$ for some fixed $\eta$, $C_{\eta}$, and $u_{\eta}$)
\end{lemma}
A number of popular kernel choices such that Gaussian, exponential, uniform, and many more satisfy the requirements of \cref{lem:jiang}. 
While the Lipschitz requirement appears strong, a large number of common distributions such as the normal distribution, Cauchy distribution, exponential distributions and lognormal distributions have Lipschitz and bounded pdfs.

\subsubsection{Balseiro-Gur Pacing Algorithm}
Consider a setting with just one episode such that the values and prices in every round $t$ are sampled from fixed stationary distributions $F$ and $D$. \citet{balseiro2019learning} give an adaptive pacing algorithm based on minimizing the dual $\Psi(\mu)$. In every round $t$, the algorithm bids $\tfrac{v_t}{1 + \mu_t}$ and the pacing parameter $\mu_t$ is updated using a projected gradient decent style update in the direction that minimizes the dual. 

\begin{lemma}[\citep{balseiro2019learning}]
\label{lem:gur}
If the value and prices in each round are samples from a stationary distribution such that $\Psi(\mu) $ is thrice differentiable in $\mu$ with bounded gradients and is strongly convex, then using the Adaptive Pacing algorithm from \citet{balseiro2019learning} with $\eta = O(T^{-1/2})$ results in strategy $\mathcal{A}$ with

\begin{align*}
  \E[]{\pp{\hind}{\v,\p}} - \E[]{\pp{\sigma}{\v ; \p}} \leq  O(\sqrt{T}).
\end{align*}

\end{lemma}

%% file: spend_pred.tex
We first turn our attention to estimating optimal spend plans in the episodic setting. Given $n$  samples from $\vec{Q}$, we will divide our budget $B$ across $E$ episodes by estimating target spend rates $(\widehat{\rho}_1, \ldots, \widehat{\rho}_E)$ that approximate the optimal spend rates $(\rho_1, \ldots, \rho_E)$ additively. The main theorem we'll prove in this section is the following.

\begin{theorem}
\label{thm:main_spend_rate_theorem}
Given $B$, $T$, $E$, sample oracles $F_e$ and $D_e$, where $d_e$ is Lipschitz and bounded, sampling budget $n$, $K$, setting  $s = O(n^{-1/3})$, w.p. $\geq 1-\frac{2E}{n}$, for each episode $e$, \approxspendrate returns $\widehat{\rho}_e$ s.t.
$$\abs{\widehat{\rho}_e - \rho_e} \leq (E+1)  \cdot \widetilde{O}\br{\frac{1}{n^{1/3}}}.$$
\end{theorem}

\approxspendrate (Algorithm \ref{alg:spend_rate}) is based on the fact that the ex-post optimal bidding strategy spends the complete budget in expectation. The resulting algorithm consists of three main steps:
i) use historical samples to approximate spend functions $\spende{\mu}$ for each episode as $\espende{\mu}$, ii) use \cref{eq:spend_all} to approximate $\mut$ as $\widehat{\mu}$, and iii) estimate the expected spend per round for each episode using the approximate spend functions and $\widehat{\mu}$.

 \begin{algorithm}
  \DontPrintSemicolon
  \SetKwInOut{Input}{Input}
  \SetKwInOut{Param}{Parameters}
  \SetKwInOut{Init}{Initialize}
  \textbf{Input:} { Budget $B$, rounds $T$, episodes $E$, sampling oracles $F_e$, $D_e$,\iffalse  number of samples $n$, \fi Kernel $K$, scalar $s$}\\

    \For{$e = 1, \ldots, E$}{
      Samples $n$ values $ \vec{V} = (V_1, V_2, \ldots, V_n) \sim F_e$\\
      Samples $n$ prices $\vec{P} = (P_1, P_2, \ldots, P_n) \sim D_e$\\
      $\espende{\mu} \leftarrow \approxspend(n, \vec{V}, \vec{P}, K, s)$
    }

  $\espend{\mu} \leftarrow  \tfrac1E\sum_{e=1}^E \espende{\mu}$
  
  $\widehat{\mu} = \min {\mu} $ s.t. $\espend{\mu} \leq \frac{B}{T}$ 
  
  \Return $(\widehat{\rho}_1, \ldots, \widehat{\rho}_E)$ where $\forall e$, $\widehat{\rho}_e  \leftarrow  \espende{\widehat{\mu}}$
  \caption{\approxspendrate}
  \label{alg:spend_rate}
\end{algorithm}

Before we discuss how to approximate the spend functions $\spende{\mu}$, in Lemma \ref{lem:spendfn_to_rate} we show that a good approximation of $\spende{\mu}$ allows for a good approximation of the optimal spend rates $\rho_1, \ldots, \rho_E$.

\begin{lemma}
In Algorithm \ref{alg:spend_rate}, for each episode $e$ if the estimated episodic spend function $\espendef$ obtained the end of Line 15 satisfies $\norm{\espendef - \spendef} \leq \gamma$, then for each $e$, the algorithm returns spend rate $\widehat{\rho}_e$ such that $\abs{\widehat{\rho}_e - \rho_e} \leq (E+1)\gamma$.
\label{lem:spendfn_to_rate}
\end{lemma}

\begin{proof}
The proof progresses in two steps: First, we show that the episodic guarantee in the premise of the lemma yields a bound for the overall spend function. Next, we show that the approximate spend functions when evaluated on $\widehat{\mut}$ yield provable bounds on the resulting episodic spend rates, where $\widehat{\mut}$ is the optimal pacing parameter learned using the estimated overall spend function.

First note that the episodic bounds yield a bound on the overall spend function $\espend$.
\begin{align}
    \abs{\espend{\mu} - \spend{\mu}} &= \abs{\frac{1}{E}\sum_1^E \espende{\mu} - \frac{1}{E}\sum_1^E \spende{\mu} } \label{eq:stof:step_1} \\
    &= \frac{1}{E}\abs{\sum_1^E(\espende{\mu} - \spende{\mu})} \nonumber\\
    &\leq \frac{1}{E} \sum_1^E \abs {\espende{\mu} - \spende{\mu}} \label{eq:stof:step_2} \\
    &\leq \frac{1}{E} \cdot E \cdot \gamma  \label{eq:stof:step_3}\\
    &= \gamma \label{eq:spend_bound}
\end{align}
Where \cref{eq:stof:step_1} follows from the definition of $\espend{\mu}$ and \cref{eq: spend_decom}, \cref{eq:stof:step_2} follow from the triangle inequality, and \cref{eq:stof:step_3} follows from the fact that $\norm{\espendef - \spendef} \leq \gamma$. Since this holds for arbitrary $\mu$, this implies that $\norm{\espendf - \spendf} \leq \gamma$.

Using \cref{eq:rho_spend_all}, and the formulation of the algorithm, we know that $\espend{\widehat{\mu}} = \spend{\mut} = \frac{B}{T}$, and \cref{eq:spend_bound} implies $\abs{\espend{\widehat{\mu}} - \spend{\widehat{\mu}}} \leq \gamma$. Combining the two, we get $\abs{\spend{\widehat{\mu}} - \spend{\mut}} \leq \gamma$

It can easily be shown that for any episode $e$, $\spende{\mu} = \E[(v, p) \sim Q_e]{\ind{v \geq (1 + \mu) p} p}$ is a monotonically decreasing function in $\mu$. Consider $\mu_1 \leq \mu_2$, since all values $v$ and prices $p$ are non-negative, for any $v$ and $p$, $\ind{v \geq (1 + \mu_2) p} p \leq \ind{v \geq (1 + \mu_1) p} p$.

So there are two possible case, 1) $\mut < \widehat{\mu}$ or $\mut \geq \widehat{\mu}$.

Consider Case 1) i.e.  $\mut < \widehat{\mu}$, then 
\begin{align}
    \abs{\spend{\widehat{\mu}} - \spend{\mut}} &= \spend{\mut} - \spend{\widehat{\mu}} \label{eq:dir_1}\\
    &= \frac{1}{E} \sum_{e=1}^E (\spende{\mut} - \spende{\widehat{\mu}}) \nonumber \\
    &\geq \frac{1}{E} \max_e(\spende{\mut} - \spende{\widehat{\mu}}) \label{eq:dir_2}
\end{align}
Where \cref{eq:dir_1} and \cref{eq:dir_2} follow from the monotonicity of $\spendf$ and $\spendef$. Similarly, using the other direction for the case $\mut \geq \widehat{\mu}$, we get that for every $e$, 
\begin{align*}
    \abs{\spende{\widehat{\mu}} - \spende{\mut}} \leq E \cdot \abs{\spend{\widehat{\mu}} - \spend{\mut}} \leq E \cdot \gamma
\end{align*}

Now consider $\abs{\widehat{\rho}_e - \rho_e}$ for some $e$, 
\begin{align*}
    \abs{\widehat{\rho}_e - \rho_e} &= \abs{\espende{\widehat{\mu}}- \spende{\mut}} \\
    &= \abs{\espende{\widehat{\mu}}- \spende{\widehat{\mu}} + \spende{\widehat{\mu}} - \spende{\mut}} \\
    &\leq \abs{\espende{\widehat{\mu}}- \spende{\widehat{\mu}}} + \abs{\spende{\widehat{\mu}} - \spende{\mut}} \\
    &\leq \gamma + E \cdot \gamma
\end{align*}
Thus, for all $e$, it holds that $\abs{\widehat{\rho}_e - \rho_e} \leq (E+1)\gamma$.
\end{proof}

\subsection{Approximating spend functions}
Recall from Definition~\ref{def:spend_function} that for an episode $e$ with value distribution $F_e$ and price distribution $D_e$, \useshortskip
\begin{align*}
 \spende{\mu} &= \textstyle \int_0^\vmax (1 - F_e((1 + \mu)p)) \cdot p \cdot d_e(p) \text{ d}p. 
\end{align*}
This implies that if we can approximate $F_e$ and $d_e$, we can use \cref{eq:spend_fn_def} to approximate $\spende{\mu}$. In \cref{alg:spend_fun}, we use the empirical  estimate $\widehat{F_e}$ of $F_e$, and use Kernel Density Estimation to approximate $d_e$ as $\widehat{d_e}$. %

{
\begin{algorithm}
  \DontPrintSemicolon
  \SetKwFunction{algo}{algo}
  \SetKwFunction{proc}{proc}
  \SetKwProg{myalg}{Protocol}{}{}
  \textbf{Input:}$(V_1, \ldots, V_n)$: values samples, $(P_1, \ldots, P_n)$: price samples, Kernel function $K$, scalar $s$\\
  
  $\widehat{d}(p) \leftarrow \frac{1}{n \cdot s} \sum_{i=1}^n K \br{\frac{p - P_i}{s}}$
  
  $\widehat{F}(v) \leftarrow \frac{1}{n} \sum_{i=1}^n \ind{V_i < v}$     
 
  $\widehat{G}(\mu) \leftarrow \int_{0}^\vmax p  (1 - \widehat{F}((1 + \mu)p))\widehat{d}(p) $ 
  
  \Return $\widehat{G}(\mu)$
  \caption{\approxspend: Stochastic prices.} \label{alg:spend_fun}
\end{algorithm} 
}

The estimate of the spend function satisfies the following uniform convergence bound.
\begin{lemma}
\label{lem:kde_spend}
Given $n$ samples from $F_e$ and $D_e$ (where $d_e$ is Lipschitz and bounded), setting $s = O(n^{-1/3})$ \approxspend (Algorithm \ref{alg:spend_fun}) returns the approximate episodic spend function $\espendef$ such that with probability at least $1 - 2/n$ it holds that
$\norm{\espendef - \spendef} \leq \widetilde{O}\br{\frac{1}{n^{1/3}}}.$ 
\end{lemma}

\begin{proof}
Consider any $\mu \geq 0$, with probability at least  $1 - \tfrac{2E}{n}$,
\begin{align*}
    \abs{\espende{\mu} - \spende{\mu}} &= \abs{\int_0^\vmax (1 - F_e((1 + \mu)p)) \cdot p \cdot d_e(p) \text{ d}p -  \int_{0}^\vmax  (1 - \widehat{F_e}((1 + \mu)p))\cdot p\cdot \widehat{d}(p)  \text{ d}p} \\
    & \leq \int_{0}^\vmax \abs{(1 - F_e((1 + \mu)p))\cdot d_e(p)- (1 - \widehat{F_e}((1 + \mu)p))\widehat{d_e}(p) }\cdot p \text{ d}p \\
    & \leq \int_{0}^\vmax \br{\abs{d_e(p) - \widehat{d_e}(p)} + \abs{\widehat{F_e}((1 + \mu)p)\widehat{d_e}(p) - F_e((1 + \mu)p)d_e(p)} } \cdot p \text{ d}p \\
    &\leq \int_{0}^\vmax \br{ \widetilde{O}\br{\frac{1}{n^{1/3}}} + \widetilde{O}\br{\frac{1}{n^{1/3}}} + \widetilde{O}\br{\frac{1}{n^{1/2}}} + \widetilde{O}\br{\frac{1}{n^{5/6}}}} \cdot p \text{ d}p \\
    & = \vmax \cdot \widetilde{O}\br{\frac{1}{n^{1/3}}} \\
    & = \widetilde{O}\br{\frac{1}{n^{1/3}}} 
\end{align*}
where the first and second inequality follow from triangle inequality. For the third step, we use the PDF and CDF concentration bounds, and the fact for any $0 \leq a,b,c,d \leq 1$, $\abs{ab - cd} \leq \abs{c-a} + \abs{d-b} + \abs{(c-a).(d-b)}$.

 Here $\widetilde{O}$ notation hides the $\textnormal{polylog}(n)$ terms along with constants like $\vmax$, $C'$ from \cref{lem:jiang} and $\norm{d_e}$.
\end{proof}

Combining the results of results of Lemma \ref{lem:spendfn_to_rate} and Lemma \ref{lem:kde_spend} completes the proof of Theorem~\ref{thm:main_spend_rate_theorem}. 
\cref{thm:main_spend_rate_theorem} implies that using $n$ historical samples, we can approximate the optimal spend rates up to an additive factor that goes down at the rate of $\Tilde{O}(n^{-1/3})$. In section \ref{app:const_price}, we show that in a simpler setting with constant prices, we can obtain a tighter error bound that goes down at the rate of $\Tilde{O}(n^{-1/2})$.

\subsection{Tighter results for Constant Prices}
\label{app:const_price}
 We consider a simpler setting where within an episode the price per impression is fixed as $p$ and only the value is sampled from distribution $F_e$. For the setting where all prices in episode $e$ are $p$, the spend function (Definition~\ref{def:spend_function}) simplifies to:
\begin{equation*}
    \spende{\mu} = (1 - F_e((1 + \mu)p)) \cdot p.
\end{equation*}
To estimate $\spende{\mu}$ we only need to estimate $F_e((1 + \mu)p)$; we give the procedure \approxspendfp in Algorithm~\ref{alg:spend_fun_fp}. The concentration guarantees for $\spende{\mu}$ follow from a straightforward application of the DKW inequality (Lemma~\ref{lemma:dkw}).

\begin{algorithm}
  \DontPrintSemicolon
  \SetKwFunction{algo}{algo}
  \SetKwFunction{proc}{proc}
  \SetKwProg{myalg}{Protocol}{}{}
  \textbf{Input:} Number of samples $n$, $(V_1, \ldots, V_n)$: values samples, $p$: price of each impression\\
  \textbf{Goal:} Estimate ${G}(\mu) = \E[(v)]{\ind{v \geq (1 + \mu) p} p}$\\
  $\widehat{F}(v) = \frac{1}{n} \sum_{i=1}^n \ind{V_n < v}$ \tcp*{Empirical cdf estimate for values}    
  $\widehat{G}(\mu) =  p(1 - \widehat{F}((1 + \mu)p))$\tcp*{Estimate spend function}
  \Return $\widehat{G}(\mu)$
  \caption{\approxspendfp: Approximate spend for constant prices.}
  \label{alg:spend_fun_fp}
\end{algorithm}

\begin{lemma}
\label{lem:fp_dkw}
Given $n$ value samples from $F_e$ and price $p$, \approxspendfp (Algorithm \ref{alg:spend_fun_fp}) returns the approximate episodic spend function $\espendef$ such that with probability at least $1 - \alpha$ it holds that
$$\norm{\espendef - \spendef} \leq p \sqrt{\tfrac{\log \tfrac{2}{\alpha}}{2n}}.$$ 
\end{lemma}
\begin{proof}
Using the DKW inequality (Lemma~\ref{lemma:dkw}), with probability at least $1-\alpha$, we have $\norm{\widehat{F_e} - F_e} \leq \sqrt{\tfrac{\log \tfrac{2}{\alpha}}{2n}}$.
Consider any $\mu \geq 0$, we have
\begin{align*}
    \abs{\espende{\mu} - \spende{\mu}} &= \abs{((1 - \widehat{F_e}(1 + \mu)p) - 1 + {F_e}((1 + \mu)p))\cdot p} \\
    & = p \cdot \abs{\widehat{F_e}((1 + \mu)p) - {F_e}((1 + \mu)p)} \\
    & \leq p \cdot \sqrt{\tfrac{\log \tfrac{2}{\alpha}}{2n}}.
\end{align*}
\end{proof}

Combining the results of results of \cref{lem:fp_dkw} and Lemma \ref{lem:spendfn_to_rate}, we can show a tighter analogue of \cref{thm:main_spend_rate_theorem} for the constant price setting.

\begin{theorem}
\label{thm:main_spend_rate_theorem_cp}
Given an episodic setting with fixed prices $p$ and parameters $B$, $T$, $E$, sampling oracles $F_e$, sampling budget $n$, $K$, with probability at least $1-\delta$, for each episode $e$, by replacing \approxspend (\cref{alg:spend_fun}) with \approxspendfp (\cref{alg:spend_fun_fp}) (at Line 7), \approxspendrate (\cref{alg:spend_rate}) returns spend rate $\widehat{\rho}_e$ such that:
\begin{align*}
    \abs{\widehat{\rho}_e - \rho_e} \leq (E+1) p \cdot \sqrt{\tfrac{\log \tfrac{2E}{\delta}}{2n}}.
\end{align*}

\end{theorem}

%% file: online_pacing.tex
Now that we have learned the spend rates, in this section we show how we can adapt the Adaptive Pacing Algorithm of \citep{balseiro2019learning}  to work with changing spend rates $\rho'_1, \cdots, \rho'_E$ which approximate the optimal spend rates.

The main idea is that using our learned spend rates, we can efficiently divide the budget across the episodes and then within each episode, we work with the budget assigned to us, and use the adaptive pacing algorithm of \citet{balseiro2019learning} as subroutine. We present this algorithm as \adaptivepacing (Algorithm \ref{alg:episodic_pace}), a detailed version of which appears as Algorithm~\ref{alg:episodic_pace_detailed} in Appendix~\ref{sec:detailed_algorithms}.

\begin{algorithm}
  \DontPrintSemicolon
  \SetKwFunction{algo}{algo}\SetKwFunction{proc}{proc}
  \SetKwProg{myalg}{Protocol}{}{}

  \textbf{Input:} Budget $B$, rounds $T$, episodes $E$, spend plan $({\rho}'_1, \ldots, {\rho}'_E)$, step size $\eta$, max shading $\bar{\mu}$.
  
  $\mu_i \leftarrow [0, \bar{\mu}]$,
  $\textsc{Budget}_{1} \leftarrow B$,
  $\te \leftarrow \frac{T}{E}$,
  $\widehat{B}_{1} \leftarrow  {\rho}'_1 \cdot \te$\\
  \For{$t=1, \ldots, T$} {
    $e \leftarrow \ceil{t/E}$ \\ 
    Observe value $v_t$\\
    Post bid $b_t \leftarrow \min \bc{ \frac{v_t}{1 + \mu_t},  \widehat{B_{e}}, \textsc{Budget}_{t} }$ \\
    Observe expenditure $z_t$\\
    $\mu_{t+1} \leftarrow \proj[0, \bar{\mu}]{\mu_t - \eta({\rho}'_e - z_t)}$ \\
    $\widehat{B}_{e} \leftarrow \widehat{B}_{e} - z_t$, 
    $\textsc{Budget}_{t+1} \leftarrow \textsc{Budget}_{t} - z_t$\\
    \If{$t\pmod E = 0$} {
      $\widehat{B}_{e+1} \leftarrow{\rho}'_{e+1} \cdot \te + \widehat{B}_{e}$ \\
    }
}    
  \caption{\adaptivepacing} \label{alg:episodic_pace}
\end{algorithm}

At the beginning of the campaign, we instantiate an overall budget \textsc{Budget} as the total budget of the campaign and an episodic budget $\widehat{B}_{1}$ for the first episode. The budget for each episode is limited ahead of time and if algorithm runs out of the episodic budget $\widehat{B}_{e}$, then it cannot buy more item in this episode, even though it may have leftover budget for the whole campaign. The intuition behind this is that the budget assigned to each episode is based on the (approximation of) the optimal spend rate. If there is left over budget after an episode ends, then the budget is simply carried forward to the next episode.

In each episode, the adaptive pacing algorithm tries to match the spend in each round to target spend rate of that round. Intuitively the algorithm works by taking the equivalent of a Stochastic Gradient Descent step in the direction of the negative of the gradient of the Lagrangian of that episode. Note that here the Lagrangian dual /average Lagrangian dual for each episode is different as is characterised by the budget for that episode. We can now show that if the spend rate estimates are good, then the resulting strategy has vanishing regret. 

\begin{definition}[Admissible Distributions]
\label{def:admissible_distributions}
Joint distribution $\vec{Q}$ such that the dual function $\duale{\mu, B_e} =  \E[(v, p) \sim Q_e]{\te\pos{v - (1 + \mu) p} +  \mu B_e}$ is thrice differentiable in $\mu$ for all $e$ and $B_e$ with bounded gradients and is strongly convex, where price distribution $D_e$ is atomic with all mass on $p$, or $d_e$ is Lipschitz and bounded.
\end{definition}

\begin{lemma}
\label{lem:final2}
If the spend rates used by Algorithm \ref{alg:episodic_pace} satisfy $\rho_e \geq \rho'_e \geq (1 - \omega) \rho_e$, with parameters $B$, $T$, $E$ resulting in strategy $\mathcal{A}$, and $\vec{Q}$ satisfies Definition~\ref{def:admissible_distributions} where $\rho_1, \cdots, \rho_E$ are the optimal spend rates, then setting $\epsilon = \omega$, we have \useshortskip
\begin{align*}
\omereg{\mathcal{A}}{T} \leq  \Tilde{O}\br{\sqrt{ET}}.
\end{align*}

\end{lemma}

To prove the lemma, we need the following additional result:
\begin{lemma}
\label{lem:rhop}
If $\rho_e \geq \rho'_e \geq (1 - \omega) \rho_e$, it holds that
$$ \inf_{\mu \geq 0 }\duale{\mu, \te \rho'_e} \geq (1 - \omega) \inf_{\mu \geq 0 }\duale{\mu, \te \rho_e}.$$
where  $\duale{\mu, B_e} = \E[(v, p) \sim Q_e]{\te\pos{v - (1 + \mu) p} +  \mu B_e}$.
\end{lemma}
\begin{proof}
Let the optimizer of $\duale{\mu, \te \rho'_e}$ be $\mu'$. We know that the optimizer of  $\duale{\mu, \te \rho_e}$ is $\mut$. Note that since $\rho_e  > \rho'_e$, using monotonicity of the spend functions, $\mut < \mu'$
Consider 
\begin{align*}
    &\duale{\mu', \te \rho'_e} - (1 - \omega) \duale{\mut, \te \rho_e} \\
    =&\te \E[(v, p) \sim Q_e]{ \pos{v - (1 + \mu') p} +   \mu'\rho'_e - (1 - \omega)\pos{v - (1 + \mut) p}  - (1 - \omega)\mut\rho_e} \\
    =& \te \E[(v, p) \sim Q_e] {(\mu'\rho'_e  - \mut(1 - \omega)\rho_e) +  \pos{v - (1 + \mu') p} - \pos{v - (1 + \mut) p} + \omega\pos{v - (1 + \mut) p}} \\
    =& \te \E[(v, p) \sim Q_e] {(\mu'\rho'_e  - \mut(1 - \omega)\rho_e) -   \ind{\mut p \leq v - p \leq  \mu'p} (v - (1 + \mut) p) + \omega\pos{v - (1 + \mut) p}} \\
    \geq &0
\end{align*}
\end{proof}

\begin{proof}[Proof of Lemma~\ref{lem:final2}]
Let's assume we divide the budget $B$ into budgets $B_e$ for all episodes $e \in [E]$. This results in an online budget constraint bid pacing problem for each individual episode. Let $\duale{\mu, B_e} = \E[(v, p) \sim Q_e]{\te\pos{v - (1 + \mu) p} +  \mu B_e}$ denote the episodic dual function for episode $e$ when the budget for episode $e$ is $B_e$. 
Similar to spend functions, if the budget allocation is optimal, that is $B_e = \tau \rho_e$, we can decompose the dual $\Psi(\mu) = \E[\v,\p]{\psi (\mu)}$ across episodes by using episodic dual functions $\duale{\mu, B_e}$.

\begin{align}
    \Psi(\mu) &= \E[\v,\p]{\br{\sum_{t=1}^T\pos{v_t - (1 + \mu) p_t}} + \mu B} \\
    &= \E[\v,\p \sim \vec{Q}] {\sum_{e=1}^E (\sum_{t = (e-1)\te + 1}^{e\te} \pos{v_t - (1 + \mu) p_t} +  \mu \rho_e )} \label{eq:ep_dual_b}\\
    &=  \te \sum_{e=1}^E \E[(v, p) \sim Q_e] {\pos{v - (1 + \mu) p} +  \mu \rho_e } \\
    &= \sum_{e=1}^E \E[(v, p) \sim Q_e]{\te\pos{v - (1 + \mu) p} +  \mu \te \rho_e} \\
    &= \sum_{e=1}^E  {\duale{\mu, \te \rho_e}} \label{eq:dual_e_rho}
\end{align}

Equation~\ref{eq:ep_dual_b} follows from Equation \ref{eq:rho_spend_all}; and Equation~\ref{eq:dual_e_rho} follows from the definition of $\duale{\mu, B_e}$. Thus $\duale{\mu, \te \rho_e}$ is the dual for the episode when the budget $B_e$ for the episode is $\te \rho_e$.

Let $\duale{\mute} = \min_{\mu \geq 0 }\duale{\mute,\te \rho_e}$. Using KKT conditions, similar to Equation \ref{eq:spend_all}, we can show that if $\mute > 0$ for all $e \in [E]$, then for all $e \in [E]$

\begin{align*}
  \dualegrad{\mute}{\te\rho_e} &= \te \rho_e - \te \spende{\mute} = 0\\
   \implies &  \rho_e =\spende{\mute} \\
   \implies & \spende{\mut} =\spende{\mute}
\end{align*}

Thus $\mut$ satisfies the KKT conditions for $\duale{\mu, \te \rho_e}$ as well. This furthermore implies that $\mut$ is an optimizer for $\duale{\mu, \te \rho_e}$. This results in the following conclusion:

\begin{align}
    \Psi(\mut) = \sum_{e=1}^E  {\duale{\mut, \te \rho_e}} = \sum_{e=1}^E  {\duale{\mute, \te \rho_e}} \label{eq:ep_dual_opt}
\end{align}

This implies that if the budget allocation across each episode is $\te\rho_e$, i.e optimal, then the optimal value of the dual can be obtained by optimizing the dual of each of the episode.

Let the strategy obtained by using our techniques be called $\mathcal{A}$. $\mathcal{A}$ uses spend rates $\rho'_e$ in each episode and assigns budget according to these rates. Once the budget has been divided, the behaviour of $\mathcal{A}$ in each episodes is independent of the other episodes. Hence we can divide the utility obtained by $\mathcal{A}$ across the episodes, i.e.
$\pp{\mathcal{A}}{\v,\p} = \sum_{e=1}^E\ppe{\mathcal{A}_e}{\v_e,\p_e}$.

Where $\ppe{\mathcal{A}_e}{\v_e,\p_e} = \sum_{t = (e-1)\te + 1}^{e\te} \bs{\ind{b^\mathcal{A}_t > p_t}(v_t - p_t)}$ and $\mathcal{A}_e$ is the strategy induced by $\mathcal{A}$ on episode $e$ by limiting the budget for $\mathcal{A}_e$ as $\rho'_e \te$.

Thus $\mathcal{A}_e$ is just the adaptive pacing strategy given in \citet{balseiro2019learning}, being run for episode $e$ with spend rate $\rho'_e$. Since things are i.i.d within the episode, we can directly use the results of \citep{balseiro2019learning}. The corresponding dual induced by the episodic sub-problem with budget $\te \rho'_e$ is $$\duale{\mu, \te \rho'_e} =  \E[(v, p) \sim Q_e]{\te\pos{v - (1 + \mu) p} +  \mu \te \rho'_e}.$$

The expected utility of $\mathcal{A}_e$ in episode $e$ is given by $\E[(v, p) \sim Q_e] {\ppe{\mathcal{A}_e}{\v_e,\p_e}}$. We use a corollary of Lemma \ref{lem:gur} which implies that by fixing the budget for episode $e$ as $\te \rho'_e$, using $\eta = O(\tau^{-1/2})$, we have
$$ \inf_{\mu \geq 0 }\ \duale{\mu, \te \rho'_e} - \ppe{\mathcal{A}_e}{\v_e,\p_e} \leq O(\sqrt{\te})$$

We know show that if the estimates $ \rho'_e $ are good, the optimal of the episodic dual with budget $ \te \rho'_e$ is not too less compared to optimal episodic dual when the budget of the episode is $ \te \rho_e$.
Using Lemma \ref{lem:rhop}, for an episode $e$
$$(1 - \omega) \inf_{\mu \geq 0 }\duale{\mu, \te \rho_e} - \E[(v, p) \sim Q_e] {\ppe{\mathcal{A}_e}{\v_e,\p_e}} = O(\sqrt{\te}).$$
Summing over all rounds and using Equation \ref{eq:ep_dual_opt} we get,
$$(1 - \omega) \inf_{\mu \geq 0 }\Psi(\mu) - \E[(\v, \p)] {\pp{\mathcal{A}}{\v,\p}} = \Tilde{O}\br{\sqrt{ET}}.$$
Using weak duality (Equation \ref{eq:weak_dual_e}), we have
$$(1 - \omega) \E[\v,\p]{\pp{\hind}{\v,\p}}  - \E[(\v, \p)] {\pp{\mathcal{A}}{\v,\p}} =  \Tilde{O}\br{\sqrt{ET}}.$$

\end{proof}

\paragraph{Putting Everything Together.}
\label{sec:end_to_end}

The final missing component is that the spend rate estimator yields an additive guarantee, while the pacing algorithm expects a multiplicative guarantee. We give a transformation for the the spend plan in Algorithm~\ref{alg:end_to_end}. Lemma~\ref{lem:add_to_mul} shows that this yields the  multiplicative guarantee.

\begin{algorithm}
  \DontPrintSemicolon
  \SetKwInOut{Input}{Input}
  \SetKwInOut{Param}{Parameters}
  \SetKwInOut{Init}{Initialize}
  \textbf{Input:} { Budget $B$, rounds $T$, episodes $E$,  sampling oracles $F_e$ and $D_e$,  per-episode sampling budget $n$, Kernel $K$, scalar $s$, step size $\eta$, max shading param $\bar{\mu}$ }\\
  $(\widehat{\rho}_1, \ldots, \widehat{\rho}_E) \leftarrow \approxspendrate(B, T, E, F_e, D_e, n, K, s)$\\
  $\widehat{\rho}'_e= \frac{(\widehat{\rho}_e + \Delta) B}{ \sum_e(\widehat{\rho}_e + \Delta) \te}$ for all $e\in 1, \ldots E$. \\
  $\adaptivepacing(B, T, E, (\widehat{\rho}'_1, \ldots, \widehat{\rho}'_E), \eta, \bar{\mu})$ 
  \caption{End-to-end algorithm}
  \label{alg:end_to_end}
\end{algorithm}

\begin{lemma}\label{lem:add_to_mul}
Given spend rates $ \widehat{\rho}_e$ such that $|\rho_e - \widehat{\rho}_e| \leq \Delta$ for all $e$, then $\widehat{\rho}'_e= \frac{(\widehat{\rho}_e + \Delta) B}{ \sum_e(\widehat{\rho}_e + \Delta) \te} \geq (1 - \frac{2 \Delta T}{B})\rho_e$ for all $e$.
\end{lemma}

\begin{proof}
From the premise it follows that
\begin{align}
\rho_e  \leq  \widehat{\rho}_e + \Delta \leq \rho_e  + 2\Delta.\label{eq:rho}
\end{align}
Similarly, scaling all sides by the constant $\tfrac{B}{ \sum_e(\widehat{\rho}_e + \Delta) \te}$, the inequalities continue to hold:
\begin{align}
 \frac{\rho_e B}{ \sum_e(\widehat{\rho}_e + \Delta) \te} \leq  \frac{\left(\widehat{\rho}_e + \Delta\right)B}{ \sum_e(\widehat{\rho}_e + \Delta) \te} \leq \frac{\left(\rho_e  + 2\Delta\right)B}{ \sum_e(\widehat{\rho}_e + \Delta) \te}.\label{eq:rho_scaled}
\end{align}
Using these observations we can derive the multiplicative lower bound:
\begin{align*}
\widehat{\rho}'_e &= \frac{(\widehat{\rho}_e + \Delta) B}{ \sum_e(\widehat{\rho}_e + \Delta) \te} \tag*{(by definition)}\\
    &\ge \frac{\rho_e B}{ \sum_e(\widehat{\rho}_e + \Delta) \te} \tag*{(by Eq.~\ref{eq:rho_scaled})}\\ 
    & \geq \frac{\rho_e B}{ \sum_e(\rho_e + 2\Delta) \te} \tag*{(since $\widehat{\rho}_e + \Delta \le \rho_e +2\Delta $ by Eq.~\ref{eq:rho})} \\
    & = \frac{\rho_e B}{ \left(\sum_e\rho_e + 2\Delta\right) \te} \\
    &=  \frac{\rho_e B}{ \left(\frac{B}{\te} + 2 \Delta\right) \te} \tag*{($\te\sum_e\rho_e = B$, Eq. \ref{eq:rho_spend_all})}\\
    &= \frac{\rho_e }{ 1 + \frac{2 \Delta T}{B}} \\
    &\geq \left(1 - \frac{2 \Delta T}{B}\right)\rho_e.
\end{align*}
\end{proof}

We can now  restate our main result formally, which follows from Lemma~\ref{lem:fp_dkw}, Lemma~\ref{lem:kde_spend}, Lemma~\ref{lem:final2}, and Lemma~\ref{lem:add_to_mul}.

\begin{theorem}[Main Theorem]
\label{thm:main_theorem_formal}
Consider the episodic setting with parameters $B$, $T$, $E$, and $n$ samples from  $\vec{Q}$ satisfying Definition~\ref{def:admissible_distributions}. Setting $s = O(n^{-1/3})$ and $\eta = O(\tau^{-1/2})$, with probability at least $1 - \delta$,  
Algorithm~\ref{alg:end_to_end} has $\omereg{\mathcal{A}}{T} \leq \Tilde{O}\br{\sqrt{ET}}$ with
\begin{itemize}
    \item $\epsilon = \frac{(E+1)pT}{B} \sqrt{\tfrac{2\log 2E/\delta}{n}}=\Tilde{O}(\frac{1}{n^{1/2}})$ for the constant-price setting, yielding vanishing regret, and
    \item $\epsilon = \Tilde{O}(\frac{1}{n^{1/3}})$ and $\delta = \frac{2E}{n}$ for the stochastic-price setting, yielding vanishing regret.
\end{itemize}
\end{theorem}

%% file: slowly_changing.tex
In this section, we consider a setting where the value and price distributions changes at every time step. In this setting, we still consider an advertiser with budget $B$ who participates in $T$ auctions. Each round $t$ has a product distribution $Q_t = F_t \times D_t$, where $F_t$ is the distribution over impression value $v_t \in \bs{0,\vmax}$ and $D_t$ over the highest competing bid $p_t \in \mathbbm{R}^+$. Thus, the $T$ round setting is characterized by distribution $\vec{Q} = (Q_1, \cdots, Q_T)$. We consider settings where this distribution changes slowly over time.

\begin{definition}[$(\zeta,\theta)-$slow-moving distribution]
A $T$ round campaign distribution $\vec{Q} = (Q_1  \cdots, Q_T)$ is called $(\zeta,\theta)-$slow moving if for all $t = 1, \cdots, T-1$, we have
\useshortskip
\begin{equation}
    \norm{F_{t+1} - F_{t}} \leq \zeta \quad\mathrm{and}\quad \norm{d_{t+1} - d_t} \leq \theta .
\end{equation}
\end{definition}

Even though the value and price distributions change in every round, since $\vec{Q}$ is \emph{slow moving}, we can generate approximately accurate spend plans by treating ranges of auctions as episodes. While the distribution in these episodes aren't stationary, the learned spend plan is approximately accurate as the distribution is slow-moving. 

\begin{algorithm}
  \DontPrintSemicolon
  \SetKwInOut{Input}{Input}
  \SetKwInOut{Param}{Parameters}
  \SetKwInOut{Init}{Initialize}
  \textbf{Input:} { Budget $B$, Total rounds $T$, Number of episodes to divide into $E$, Sampling oracles $F_t$ for values and $D_t$ for prices, sampling budget $n$, Kernel $K$, scalar $s$, step size $\eta$, max shading param $\bar{\mu}$ }\\

Divide $T$ into $E$ episodes of size $\tau = \frac{T}{E}$

Construct episodic sampling oracles $\widetilde{F_e} = \frac{1}{\tau} \sum_{t = e.\tau + 1}^{(e+1)\tau} F_t$ and $ \widetilde{D_e} = \frac{1}{\tau} \sum_{t = e.\tau + 1}^{(e+1)\tau} D_t$

  $(\widehat{\rho}_1, \ldots, \widehat{\rho}_E) \leftarrow \approxspendrate(B, T,  E, \widetilde{F_e} ,  \widetilde{D_e} , n, K, s)$\\
  $\widehat{\rho}'_e= \frac{(\widehat{\rho}_e + \Delta) B}{ \sum_e(\widehat{\rho}_e + \Delta) \te}$ for all $e\in 1, \ldots E$. \\
  $\adaptivepacing(B, T, E, (\widehat{\rho}'_1, \ldots, \widehat{\rho}'_E), \eta, \bar{\mu})$ 
  \caption{Spend Prediction and Pacing for Slowly Changing Distribution}
  \label{alg:end_to_end_slowly}
\end{algorithm}

\begin{definition}[Admissible Moving Distributions]
\label{def:admissible_distributions_slowly}
Joint distribution $\vec{Q}$ s.t. it satisfies definition \ref{def:admissible_distributions} and for any rounds $i$ and $j$ that fall in the same episode, the spend function is strongly monotone, i.e. $(\mu' -\mu)(\overline{G_i}(\mu) - \overline{G_j}(\mu')) > C (\mu' - \mu)^2$ for some constant $C$.
\end{definition}

As per definition \ref{def:spend_function}, the average spend function for rounds in episode $e$ can be given as $ \spende{\mu} = \frac{1}{\tau} \sum_{t = e.\tau + 1}^{(e+1)\tau}\E[(v, p) \sim Q_t]{\ind{v \geq (1 + \mu) p} p}$ and for the ex-post optimal bidding strategy of bidding $\frac{v_t}{1 + \mut}$, we have 
\begin{equation}
   \tau \sum_{e=1}^E \spende{\mut} =  \tau \sum_{e=1}^E \rho_e = B.
   \label{eq:spend_all_slowly}
\end{equation}
where $\rho_1,\cdots, \rho_E$ are the optimal spend rates. The accuracy of the spend plan now depends on the choice for $E$ and parameters $\zeta$ and $\theta$ that capture how fast the distribution is changing. 
\begin{lemma}
\label{thm:main_spend_rate_theorem_slow}
Given $B$, $T$, sampling oracles $F_t$ and $D_t$ such that $\vec{Q}$ is $(\zeta,\theta)-$slow moving, number of episodes to break into $E$, $n$, $K$, then with probability at least $1-\frac{2E}{n}$, using \cref{alg:end_to_end_slowly}, in Line 4, \approxspendrate (\cref{alg:spend_rate}) returns spend rates $\widehat{\rho}_e$ such that for every episode $e$, 
$$\abs{\widehat{\rho}_e - \rho_e} \leq (E+1)  \cdot \widetilde{O}\br{\frac{1}{n^{1/3}} + \frac{T(\zeta + \theta)}{E}}$$provided $d_t$ is Lipschitz and bounded by setting $s = n^{-1/3}$.
\end{lemma}

\begin{proof}
Consider $\widetilde{F_e} = \frac{1}{\tau} \sum_{t = e.\tau + 1}^{(e+1)\tau} F_t$. For any round $t$ belonging to episode $e$, we have
\begin{align}
    \norm{F_t - \widetilde{F_e}} &=  \frac{1}{\tau} \norm{\tau.F_t - \sum_{x = \floor{t/\tau}*\tau + 1}^{(\floor{t/\tau}+1)*\tau}{F_x}} \leq \frac{1}{\tau} \sum_{x = \floor{t/\tau}*\tau + 1}^{(\floor{t/\tau}+1)*\tau}  \norm{F_x - F_t} \leq \frac{\zeta \tau}{2}.
    \label{eq:avg_cdf}
\end{align}
Similarly, for any round $t$ belonging to episode $e$ we have
\begin{equation}
    \norm{d_t - \widetilde{d_e}} \leq  \frac{\theta \tau}{2}.
     \label{eq:avg_pdf}
\end{equation}

Let $\widehat{F_e}$ be the empirical cdf obtained using $n$ samples from $\widetilde{F_e}$. Using the DKW inequality (Lemma~\ref{lemma:dkw}), with probability at least $1-\alpha$, we have $\norm{\widehat{F_e} - \widetilde{F_e}} \leq \sqrt{\tfrac{\log \tfrac{2}{\alpha}}{2n}}$. Similarly, let $\widehat{d_e}$ be the kernel density estimate of  $\widetilde{d_e}$ obtained using $n$ samples. Using Lemma \ref{lem:jiang}, with probability at least $1 - \frac{1}{n}$, we have $\norm{\widehat{d_e} - \widetilde{d_e}} =  \widetilde{O}\br{\frac{1}{n^{1/3}}}$. Combining the concentration results with \cref{eq:avg_cdf} and \cref{eq:avg_pdf}, we that that with probability at least $1 - \frac{2E}{n}$, for all episodes $e$
\begin{equation}
    \norm{F_t - \widehat{F_e}} \leq  \sqrt{\tfrac{\log \tfrac{n}{E}}{2n}} + \frac{\zeta \tau}{2} \quad \mathrm{and} \quad  \norm{d_t - \widehat{d_e}} \leq   \widetilde{O}\br{\frac{1}{n^{1/3}}} + \frac{\theta \tau}{2}.
    \label{eq:slow_close_estimates}
\end{equation}
Consider the episodic spend function induced by $ \widehat{F_e}$ and $\widehat{d_e}$ as $\espende{\mu}$. For all episodes $e$, with probability at least $1 - \frac{2E}{n}$, we have

\begin{align*}
    \abs{\spende{\mu} - \espende{\mu}} &= \abs{\frac{1}{\tau} \sum_{t = e.\tau + 1}^{(e+1)\tau} \int_0^\vmax (1 - F_t((1 + \mu)p)) \cdot p \cdot d_t(p) \text{ d}p -  \int_{0}^\vmax  (1 - \widehat{F_e}((1 + \mu)p))\cdot p\cdot \widehat{d_e}(p)  \text{ d}p} \\
    & \leq \frac{1}{\tau} \sum_{t = e.\tau + 1}^{(e+1)\tau} \abs{ \int_0^\vmax (1 - F_t((1 + \mu)p)) \cdot p \cdot d_t(p) \text{ d}p - \int_{0}^\vmax  (1 - \widehat{F_e}((1 + \mu)p))\cdot p\cdot \widehat{d_e}(p)  \text{ d}p} \\
    & \leq \frac{1}{\tau} \sum_{t = e.\tau + 1}^{(e+1)\tau} \int_{0}^\vmax \abs{(1 - F_t((1 + \mu)p))\cdot d_t(p)- (1 - \widehat{F_e}((1 + \mu)p))\widehat{d_e}(p) }\cdot p \text{ d}p \\
    & \leq \frac{1}{\tau} \sum_{t = e.\tau + 1}^{(e+1)\tau}  \int_{0}^\vmax \br{\abs{d_t(p) - \widehat{d_e}(p)} + \abs{\widehat{F_e}((1 + \mu)p)\widehat{d_e}(p) - F_t((1 + \mu)p)d_t(p)} } \cdot p \text{ d}p \\
    & \leq \frac{1}{\tau} \sum_{t = e.\tau + 1}^{(e+1)\tau} \int_{0}^\vmax \br{  \widetilde{O}\br{\frac{1}{n^{1/3}}} + \frac{\theta \tau}{2}  + \widetilde{O}\br{\frac{1}{n^{1/3}}} + \frac{\theta \tau}{2}} \cdot p \text{ d}p \\
    & + \frac{1}{\tau} \sum_{t = e.\tau + 1}^{(e+1)\tau} \int_{0}^\vmax \br{   \sqrt{\tfrac{\log \tfrac{n}{E}}{2n}} + \frac{\zeta \tau}{2} + \widetilde{O}\br{\frac{\zeta \tau }{n^{1/3}}} + \widetilde{O}\br{\frac{\theta \tau }{n^{1/2}}} + \widetilde{O}\br{\frac{1}{n^{5/6}}} } \cdot p \text{ d}p \\
    & = \vmax \cdot \widetilde{O}\br{\frac{\zeta \tau + 1}{n^{1/3}} + \zeta \tau + \theta \tau} \\
    & = \widetilde{O}\br{\frac{\zeta \tau + 1}{n^{1/3}} + \zeta \tau + \theta \tau} 
\end{align*}
where the first, second, and third inequality follow from triangle inequality. For the fourth step, we use \cref{eq:slow_close_estimates} and the fact for any $0 \leq a,b,c,d \leq 1$, $\abs{ab - cd} \leq \abs{c-a} + \abs{d-b} + \abs{(c-a).(d-b)}$.
Using Lemma \ref{lem:spendfn_to_rate}, we get that for all $e$, with probability at least $1 - \frac{2e}{n}$, we have  $\abs{\widehat{\rho}_e - \rho_e} \leq (E+1)  \cdot \widetilde{O}\br{\frac{\zeta \tau}{n^{1/3}} + \frac{\tau(\zeta + \theta)}{E}}$. 

\end{proof}

Now that we have approximate spend rates, we can use the estimates to divide the budget across the smaller episodes and use $\adaptivepacing$ (Algorithm \ref{alg:episodic_pace}) to perform online pacing. Combining all the guarantees, we can show the following main result for this setting.

\begin{theorem}
For the pacing setting with parameters $B$, $T$, $\vec{Q}$, number of episodes to break into $E$, Kernel $K$, if $\vec{Q}$ is $(\zeta,\theta)-$slow moving and satisfies Definition \ref{def:admissible_distributions_slowly}, given $n$ samples from $\vec{Q}$, by setting $s = n^{-1/3}$ and $\eta = \tau^{-1/2}$, with probability at least $1 - \frac{2E}{n}$, Algorithm \ref{alg:end_to_end_slowly} resulting in strategy $\mathcal{A}$ has $\omereg{\mathcal{A}}{T} \leq \Tilde{O}\br{\sqrt{ET}}$ with $\epsilon = \frac{2ET}{B}  \cdot \widetilde{O}\br{\frac{1}{n^{1/3}} + \frac{T(\zeta + \theta)}{E}}$.
\label{thm:end_to_end_slow}
\end{theorem}

Theorem \ref{thm:end_to_end_slow} is implied by  combining \cref{thm:main_spend_rate_theorem_slow},  \cref{lem:add_to_mul}, and \cref{lem:final2}.
Theorem \ref{thm:end_to_end_slow} implies our results for the episodic setting can be extended to obtain results for more general settings. We can also observe that in this case, the $\epsilon$ in $\omereg{\mathcal{A}}{T}$ doesn't converge to $0$ as $n$ grows, since the nonstationarity  within an episode  does not decrease with more samples.

%% file: experiments.tex
In this section we give empirical evidence that our end-to-end pipeline of optimal spend rate estimation and online pacing provides noticeable benefits over episode-blind pacing schedules and truthful bidding in a variety of synthetic environments.

\paragraph{Datasets.} We create synthetic datasets to test the performance of the algorithms under consideration. For the values, we consider three distributions: uniform, normal, and lognormal. For the prices, we consider three settings: fixed prices (our analysis focuses on this setting before generalizing), normally distributed prices, and the max of multiple draws from a lognormal distribution.\footnote{Prior work, e.g. \cite{ostrovsky2011reserve}, suggest that bids in ad auctions typically follow a lognormal distribution. The combination of lognormal values with max-of-lognormal-draws as prices is a realistic simulation of auction environment which is captured in the lognorn\_v\_maxlognorm\_p  dataset.}  We combine these into $6$ synthetic datasets, see Table~\ref{tab:datasets}.  We divide the time horizon into 10 episodes, i.e. $E=10$ with differing parameters of distributions for each episode and use $T = 1000$.

\begin{table}[t]
\caption{Descriptions of synthetic datasets.}
\label{tab:datasets}
\begin{center}
\begin{tabular}{lcc}
\toprule
Data set & Value distributions $F_e$ & Price distributions $D_e$ \\
\midrule
uniform\_v\_fix\_p   & Uniform dist over $[l_e, r_e]$ & Fixed price $p$ \\
normal\_v\_fix\_p   & $\mathcal{N}(\mu_{v,e}, \sigma^2_{v,e})$ & Fixed price $p$ \\
lognorm\_v\_fix\_p   & $\textnormal{Lognormal}(\mu_{v,e}, \sigma^2_{v,e})$ & Fixed price $p$ \\
uniform\_v\_normal\_p   & Uniform dist over $[l_e, r_e]$ & $\mathcal{N}(\mu_{p,e}, \sigma^2_{p,e})$ \\
normal\_v\_normal\_p   & $\mathcal{N}(\mu_{v,e}, \sigma^2_{v,e})$& $\mathcal{N}(\mu_{p,e}, \sigma^2_{p,e})$ \\
lognorn\_v\_maxlognorm\_p   & $\textnormal{Lognormal}(\mu_{v,e}, \sigma^2_{v,e})$ & $\max_{k \in [K]}\textnormal{Lognormal}(\mu_{k,e}, \sigma^2_{k,e})$ \\
\bottomrule
\end{tabular}
\end{center}
\end{table}

\subsection{Learning Accurate Spend Plans}
\begin{figure*}
	\centering\includegraphics[width=0.8\linewidth]{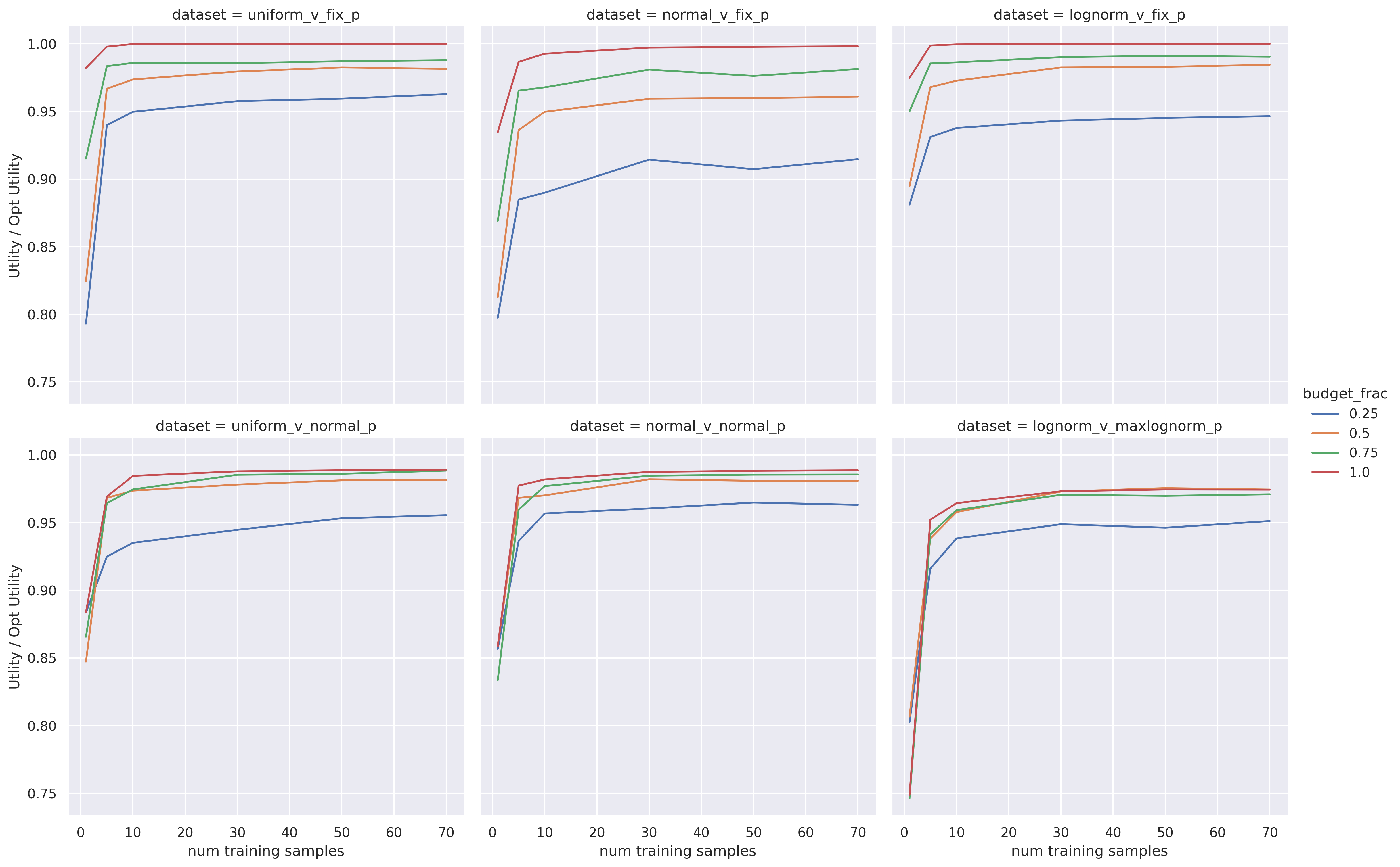}
	\caption{\small Performance of the end-to-end pacing system as a function of the size of training data. We plot the ratio of the optimal utility that our pacing system is able to obtain as a function of the number of training samples in the rate rate estimation phase. The budget is represented as budget\_frac, i.e. $B =$ budget\_frac*$\bar{C}$. We can see that 1) with increasing samples, the performance improves quickly 2) the budget level is important for the overall performance.}
	\label{fig:sample_complexity}
\end{figure*}

 To understand how the performance of our end-to-end pacing is dependent on the numbers of samples available in the spend plan estimation phase, we plot the ratio of the optimal utility that the pacing system is able to obtain as a function of the number of training samples in Figure \ref{fig:sample_complexity}. We consider 4 different budget levels: let $\bar{C}$ be the expenditure of the campaign that bids their value in each auction, we consider budgets $x\bar{C}$ for $x \in \{0.25, 0.5, 0.75, 1.0 \}$.

Figure~\ref{fig:sample_complexity} shows the effect of varying the training set size, where a sample in the training set corresponds to a value and price draw from each episode. Two things are clear from the results: 1) with increasing samples, the performance improves quickly, and 2) the budget level is important for the overall performance as the optimal budget allocation problem gets harder for smaller budgets. This is in agreement with the theoretical results of the pacing system obtained in Theorem \cref{thm:main_theorem_formal}. Finally, it does appear that the performance hits a plateau. This is likely due to the online part of the algorithm which does not scale with increased offline learning sample size.

\begin{figure*}[]
\centering	\includegraphics[width=0.9\linewidth]{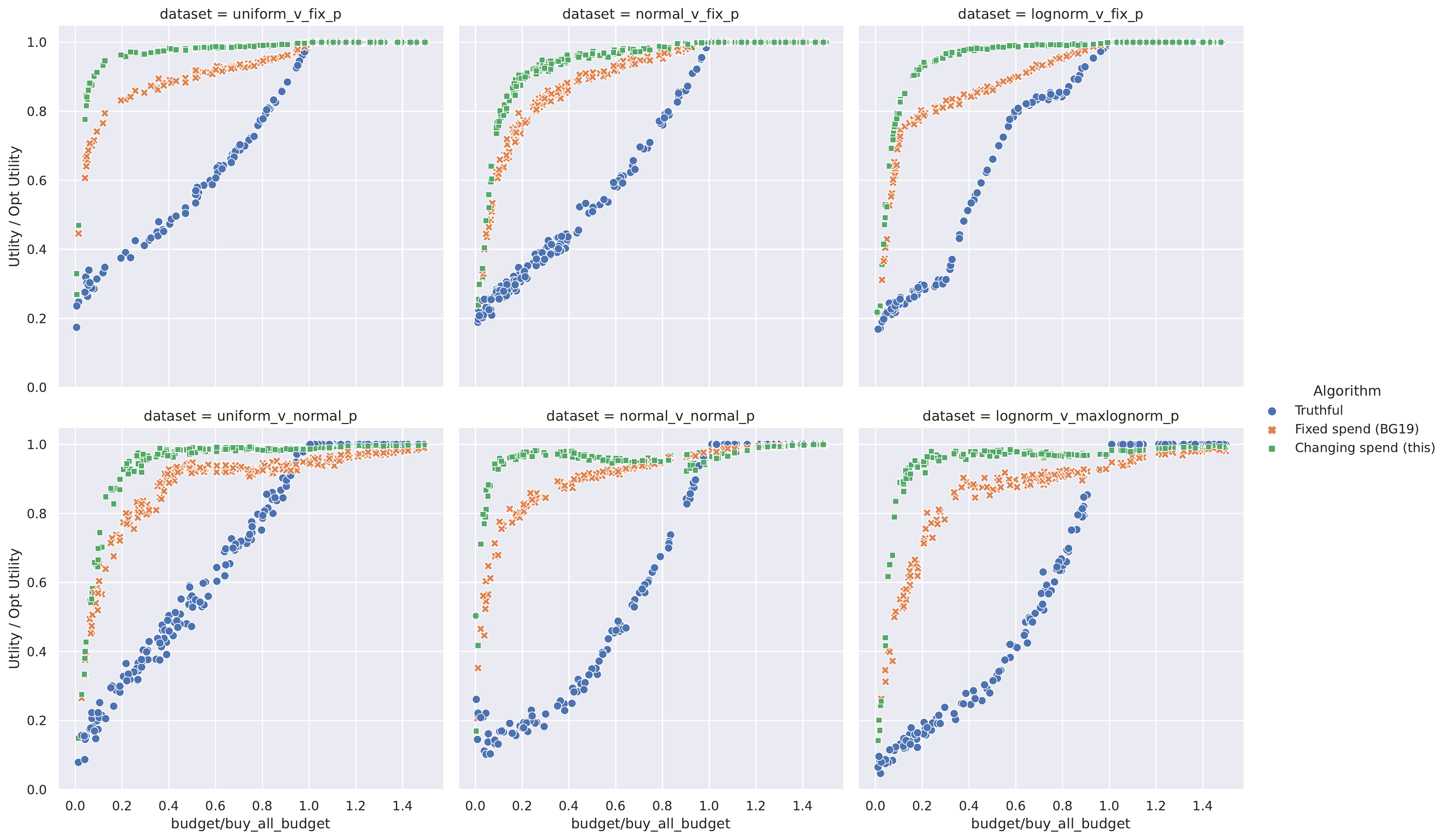}
	\caption{ \small Comparing performance of our algorithm labeled as \emph{Changing spend (this)}, fixed spend rate \cite{balseiro2019learning} labeled as \emph{Fixed spend (BG19)} and no pacing labeled as \emph{Truthful} on synthetic datasets. See Table \ref{tab:datasets} for details on datasets. Each datapoint in the scatter plot refers to one experiment where we plot the fraction of the optimal utility obtained by the pacing strategy as a function of the budget \emph{buy\_all\_budget} represents $\bar{C}$. In each of these cases, our method achieves a higher fraction of optimal utility than either no pacing (truthful bidding) or fixed spend rate pacing strategies (\cite{balseiro2019learning}) over nearly all ratios of the budget relative to the cost of all impressions. }
	\label{fig:toy_data}
\end{figure*}

\subsection{End-to-end Performance}\label{sec:exp_synth}
{\bf Algorithms.}
We compare the performance of 1) \emph{our algorithm}, 2) \emph{constant spend rate} \citep{balseiro2019learning} (which uses a linear cumulative expenditure over the duration of the campaign, and  gives us an understanding of the benefit of estimating spend rates when competition is time-varying), and 3) \emph{no pacing} (which enters the advertiser's value in each auction until they run out of budget).
 
{\bf Results.} The values and prices were generated in the same way as above. For each dataset, we run simulations where we a budget for the campaign is drawn uniformly from $[0, \bar{C}]$, where $\bar{C}$ is the expenditure of the campaign that bids truthfully in each auction. Then we run all the pacing algorithms for this dataset sample and budget level. We repeat this process 150 times to get 150 data points per dataset for each algorithm. We plot the ratio of the optimal utility that the pacing system is able to obtain as a function of the budget level in Figure \ref{fig:toy_data}. Our algorithm outperforms both benchmarks almost everywhere. The only time where the ``Truthful'' benchmark performs better are in situations where the advertiser has enough budget to buy (almost) all impressions. There is one area where ``Fixed spend (BG19)'' outperforms our algorithm. It happens for the ``normal\_v\_normal\_p'' dataset when $B\ge 0.8\cdot \bar{C}$; we do not have an explanation why this particular range performs poorly.

%% file: appendix.tex
 \section{Characterizing the optimal pacing strategy and budget allocation in expectation}
 \label{app:optimal_b}
Recalling that the optimal strategy on the realized values and prices is obtained by the hindsight strategy $H$ (definition~\ref{def:benchmark}). The Lagrangian dual of the optimization problem in definition~\ref{def:benchmark} is given by: 

\begin{align}
\label{prob:dual}
  \begin{split}
    \psi (\mu) =  \bs{\sum_{t=1}^T\pos{v_t - (1 + \mu) p_t}} + \mu B \\
\end{split}
&& \text{(L(KP))} 
\end{align}
Where we define $\pos{z}$ to be $\max\bc{z,0}$. The dual is obtained from the Lagrangian by setting $x_t = 1$ for all $t$ such that $v_t - (1 + \mu)p_t \geq 0$, i.e. winning all impressions with value greater than $(1 + \mu)$ times the price which can be done by bidding $b_t = v_t/(1 + \mu)$.

By weak duality, we have 
\begin{equation}
\label{eq:weak_dual}
   \pp{\hind}{{\v , \p}} \leq  \inf_{\mu \geq 0} \psi (\mu) 
\end{equation}

Since $\v$  and $\p$ are being sampled from the fixed distribution defined by $\vec{Q}$, taking expectation over equation \ref{eq:weak_dual} and using Jensen's inequality, we get

\begin{equation}
\label{eq:weak_dual_e}
   {\E[\v,\p]{\pp{\hind}{\v,\p}}} \leq  \E[\v,\p]{\inf_{\mu \geq 0} \psi (\mu)} \leq  \inf_{\mu \geq 0}  \E[\v,\p]{\psi (\mu)}
\end{equation}

Let $\Psi(\mu) = \E[\v,\p]{\psi (\mu)}$ and $\mut$ be the minimizer of $\Psi(\mu)$. Assuming $\Psi(\mu)$ to be differentiable, using Karush-Kuhn-Tucker conditions, we have $\mut \geq 0$, $\Psi'(\mut) \geq 0$, and $\mut \Psi'(\mut) = 0$. If $\mut = 0$, it implies that we are effectively not constrained by the budget and truthful bidding achieves the optimal utility in expectation as it wins all items with positive utility. 

The gradient of $\Psi(\mu)$ can be written as $$\Psi'(\mu) = B - G(\mu)$$ where $$G(\mu) = \E[\v,\p]{\sum_{t=1}^T \ind{v_t \geq (1 + \mu) p_t} p_t}$$

We call $G(\mu)$ the overall spend function. By definition, $G(\mu)$ is the expected expenditure over all the $T$ rounds when buying all items such that $v_t \geq (1+\mu) p_t$, obtained by bidding $b_t = v_t/(1 + \mu)$. The KKT complementary slackness condition implies that if $\mut > 0$, then $\Psi'(\mut) = 0$ i.e.
\begin{equation}
\label{eq:spend_all_app}
    G(\mut) = B
\end{equation}

This implies that the strategy with a fixed pacing multiplier that bids $b_t = v_t/(1 + \mut)$ achieves better expected utility than the expected utility of the hindsight strategy $H$. If truthful bidding is not optimal (i.e $\mut > 0$ ), then the expected expenditure of this strategy is $B$. Not that the expenditure guarantee for the optimal fixed shading strategy is only satisfied in expectation, i.e. it spends budget $B$ in expectation.

For the rest of the theoretical claims, we restrict ourselves to the case that $\mut > 0$. The case $\mut = 0$ implies that the budget constraint is not binding, so truthful bidding is the optimal strategy. Our algorithm will naturally adapt to this setting as well.

Let $\spend{\mu} = \frac{G(\mu)}{T}$ be the average spend in each round (over the whole campaign) if we buy all items such that $v_t \geq (1+\mu) p_t$. Equation \ref{eq:spend_all_app} implies that for the optimal dual variable $\mut > 0,$
\begin{equation}
\label{eq:spend_all_ave}
    \spend{\mut} = \frac{B}{T}.
\end{equation}

Using Definition \ref{def:spend_function}, $\spende{\mu} = \E[(v, p) \sim Q_e]{\ind{v \geq (1 + \mu) p} p}$. $\spende{\mu}$ is the expected spend per round in an episode $e$ if the dual variable is $\mu$, corresponding to the strategy which bids by multiplicatively shading the value $v_t$ by a factor of $\frac{1}{1 + \mu}$ and ends up buying all the impressions in the episode with value per unit spent at least $(1+\mu)$.
In our framework, the spend function can be decomposed across the episodes by introducing  episodic spend functions. We show this decomposition below:

\begin{align}
\begin{split}
    G(\mu) &=  \E[\v,\p \sim \vec{Q}]{\sum_{t=1}^T \ind{v_t \geq (1 + \mu) p_t} p_t} \\
    &= \E[\v,\p \sim \vec{Q}] {\sum_{e=1}^E \sum_{t = (e-1)\te + 1}^{e\te} \ind{v_t \geq (1 + \mu) p_t} p_t} \\
    &= \sum_{e=1}^E \sum_{t = (e-1)\te + 1}^{e\te} \E[(v_t, p_t) \sim Q_e]{\ind{v_t \geq (1 + \mu) p_t} p_t} \\
    &= \te \sum_{e=1}^E \E[(v, p) \sim Q_e]{\ind{v \geq (1 + \mu) p} p} \\
    &= \te \sum_{e=1}^E \spende{\mu} \\
\implies \frac{G(\mu)}{T} &= \frac{1}{E}\sum_{e=1}^E \spende{\mu} \\
\implies \spend{\mu}  &= \frac{1}{E}\sum_{e=1}^E \spende{\mu}
\end{split}\label{eq: spend_decom}
\end{align}
where $\spende{\mu} = \E[(v, p) \sim Q_e]{\ind{v \geq (1 + \mu) p} p}$  is the episodic spend function. 
This definition helps us to define the optimal budget allocation as $B_e = \tau \spende{\mut} = \tau \rho_e$ where $\rho_e$ is the \emph{optimal spend rate} for episode $e$ given by $\rho_e =  \spende{\mut} $. Note that if $\mut >0$, using Equation \ref{eq:spend_all}, we have
\begin{equation}
\label{eq:rho_spend_all}
    \te \sum_{e=1}^E \rho_e = \te\sum_{e=1}^E \spende{\mut} = G(\mut) = B
\end{equation}

\section{Detailed Algorithms}
\label{sec:detailed_algorithms}

\subsection{\adaptivepacing: Adaptive pacing using a spend plan}
\begin{algorithm}[]
  \DontPrintSemicolon
  \SetKwFunction{algo}{algo}\SetKwFunction{proc}{proc}
  \SetKwProg{myalg}{Protocol}{}{}

  \textbf{Input:} Budget $B$, rounds $T$, episodes $E$, spend plan $({\rho}'_1, \ldots, {\rho}'_E)$, step size $\eta$, max shading param $\bar{\mu}$ 
  
  $\mu_i \leftarrow [0, \bar{\mu}]$  \tcp*{Initialize shading multiplier}
  $\textsc{Budget}_{1} \leftarrow B$ \tcp*{Overall remaining budget left for campaign}
  $\te \leftarrow \frac{T}{E}$ \tcp*{Impressions in each episode}
  $\widehat{B}_{1} \leftarrow  {\rho}'_1 \cdot \te$ \tcp*{Remaining budget for episode 1}
  \For{$t=1, \ldots, T$} {
    $e \leftarrow \ceil{t/E}$ \tcp*{Current episode} 
    Observe value $v_t$\\
    Post bid $b_t \leftarrow \min \bc{ \frac{v_t}{1 + \mu_t},  \widehat{B_{e}}, \textsc{Budget}_{t} }$ \\
    Observe expenditure $z_t$\\
    $\mu_{t+1} \leftarrow \proj[0, \bar{\mu}]{\mu_t - \eta({\rho}'_e - z_t)}$ \tcp*{Update shading parameter} 
    $\widehat{B}_{e} \leftarrow \widehat{B}_{e} - z_t$\tcp*{Update remaining budget}
    $\textsc{Budget}_{t+1} \leftarrow \textsc{Budget}_{t} - z_t$\\
    \If{$t\pmod E = 0$} {
      $\widehat{B}_{e+1} \leftarrow{\rho}'_{e+1} \cdot \te + \widehat{B}_{e}$ \tcp*{Carry over left-over budget}
    }
}    
  \caption{\adaptivepacing: Adaptive pacing using a spend plan.} \label{alg:episodic_pace_detailed}
\end{algorithm} 

We present \approxspendrate  (Algorithm~\ref{alg:spend_rate}) which uses historical data to compute approximately optimal spend rates $(\widehat{\rho}_1, \ldots, \widehat{\rho}_E)$ from samples.

For each episode $e$, the subroutine $\approxspend$ (Algorithm \ref{alg:spend_fun}) estimates the episodic spend function $\spende{\mu}$ as a function of $\mu$ using the historic samples $\vec{V}$ and $\vec{P}$. For fixed prices, we use a simpler episodic spend prediction function estimate $\approxspendfp$ (Algorithm \ref{alg:spend_fun_fp}). Both functions try to estimate $\spende{\mu}$ and return an empirical approximate of  $\spende{\mu}$ we denote as $\espende{\mu}$.

Then using the structure of overall average spend function $\spend{\mu}$, (Equation \ref{eq: spend_decom}), we can construct an approximation of the overall average spend function $\espend{\mu}$ as $\frac{\sum_{e=1}^E \espende{\mu}}{E}$. Based on our discussion about the optimal structure of the problem, for optimal dual variable $\mut$, we know that $\spend{\mut} = \frac{B}{T}$ (Equation \ref{eq:spend_all}). Using our empirical estimate  $\espend{\mu}$, we compute $\widehat{\mu}$, an empirical estimate of $\mut$. We can compose our approximations to form $\widehat{\rho}_e = \espende{\widehat{\mu}}$, an approximation to $\spende{\mut} = \rho_e$.

\begin{algorithm}[]
  \DontPrintSemicolon
  \SetKwInOut{Input}{Input}
  \SetKwInOut{Param}{Parameters}
  \SetKwInOut{Init}{Initialize}
  \textbf{Input:} { Budget $B$, Total rounds $T$, Number of episodes $E$, Episodic sampling oracles $F_e$ for values and $D_e$ for prices, Per episode sampling budget $n$, Kernel $K$, scalar $s$}\\
  \textbf{Goal:} Estimate optimal spend rates $(\rho_1, \ldots, \rho_E)$\\
 
  \If{in the constant-price setting}{
    \For{$e = 1, \ldots, E$}{
      Samples n values $ \vec{V} = (V_1, V_2, \ldots, V_n) \sim F_e$\\
      Set price $p \sim D_e$\\
      $\espende{\mu} = \approxspendfp(n, \vec{V}, p)$ \tcp*{Estimate episodic spend function} 
    } 
  } \Else {
    \For{$e = 1, \ldots, E$}{
      Samples n values $ \vec{V} = (V_1, V_2, \ldots, V_n) \sim F_e$\\
      Samples n prices $\vec{P} = (P_1, P_2, \ldots, P_n) \sim D_e$\\
      $\espend{\mu} = \approxspend(n, \vec{V}, \vec{P}, K, s)$\tcp*{Estimate episodic spend function}
    }
  }
  $\spende{\mu} =  \frac{\sum_{e=1}^E \espende{\mu}}{E}$ \tcp*{Construct overall average spend function} 
  $\widehat{\mu} = \min {\mu} $ s.t. $\spende{\mu} \leq \frac{B}{T}$ \tcp*{Estimating the optimal dual variable} 
  \For{$e = 1, \ldots, E$}{
    $\widehat{\rho}_e = \espende{\widehat{\mu}}$ \tcp*{Expected spend rate in episode for the estimated dual variable}
  }
  \Return $(\widehat{\rho}_1, \ldots, \widehat{\rho}_E)$
  \caption{\approxspendrate: Approximate optimal spend rates}
  \label{alg:spend_rate_delated}
\end{algorithm}

%% file: main_paper.bbl
\begin{thebibliography}{38}
\providecommand{\natexlab}[1]{#1}
\providecommand{\url}[1]{\texttt{#1}}
\expandafter\ifx\csname urlstyle\endcsname\relax
  \providecommand{\doi}[1]{doi: #1}\else
  \providecommand{\doi}{doi: \begingroup \urlstyle{rm}\Url}\fi

\bibitem[Abrams et~al.(2008)Abrams, Keerthi, Mendelevitch, and
  Tomlin]{abrams2008comprehensive}
Zo{\"o} Abrams, S~Sathiya Keerthi, Ofer Mendelevitch, and John~A Tomlin.
\newblock Ad delivery with budgeted advertisers: A comprehensive lp approach.
\newblock \emph{Journal of Electronic Commerce Research}, 9\penalty0 (1), 2008.

\bibitem[Agarwal et~al.(2010)Agarwal, Chen, Lin, Shanmugasundaram, and
  Vee]{agarwal2010forecasting}
Deepak Agarwal, Datong Chen, Long-ji Lin, Jayavel Shanmugasundaram, and Erik
  Vee.
\newblock Forecasting high-dimensional data.
\newblock In \emph{Proceedings of the 2010 ACM SIGMOD International Conference
  on Management of data}, pages 1003--1012, 2010.

\bibitem[Agarwal et~al.(2014)Agarwal, Ghosh, Wei, and You]{agarwal2014budget}
Deepak Agarwal, Souvik Ghosh, Kai Wei, and Siyu You.
\newblock Budget pacing for targeted online advertisements at linkedin.
\newblock In \emph{Proceedings of the 20th ACM SIGKDD international conference
  on Knowledge discovery and data mining}, pages 1613--1619, 2014.

\bibitem[Amin et~al.(2012)Amin, Kearns, Key, and Schwaighofer]{amin2012budget}
Kareem Amin, Michael Kearns, Peter Key, and Anton Schwaighofer.
\newblock Budget optimization for sponsored search: Censored learning in mdps.
\newblock \emph{arXiv preprint arXiv:1210.4847}, 2012.

\bibitem[Avadhanula et~al.(2020)Avadhanula, Colini-Baldeschi, Leonardi,
  Abinav~Sankararaman, and Schrijvers]{avadhanula2020}
Vashist Avadhanula, Riccardo Colini-Baldeschi, Stefano Leonardi, Karthik
  Abinav~Sankararaman, and Okke Schrijvers.
\newblock Stochastic bandits for multi-platform budget optimization in online
  advertising.
\newblock In \emph{The World Wide Web Conference}, 2020.

\bibitem[Azar et~al.(2009)Azar, Birnbaum, Karlin, and Nguyen]{azar2009gsp}
Yossi Azar, Benjamin Birnbaum, Anna~R. Karlin, and C.~Thach Nguyen.
\newblock On revenue maximization in second-price ad auctions.
\newblock In Amos Fiat and Peter Sanders, editors, \emph{Algorithms - ESA
  2009}, pages 155--166, Berlin, Heidelberg, 2009. Springer Berlin Heidelberg.
\newblock ISBN 978-3-642-04128-0.

\bibitem[Babaioff et~al.(2020)Babaioff, Cole, Hartline, Immorlica, and
  Lucier]{babaioff2020non}
Moshe Babaioff, Richard Cole, Jason Hartline, Nicole Immorlica, and Brendan
  Lucier.
\newblock Non-quasi-linear agents in quasi-linear mechanisms.
\newblock \emph{arXiv preprint arXiv:2012.02893}, 2020.

\bibitem[Balseiro et~al.(2017)Balseiro, Kim, Mahdian, and
  Mirrokni]{balseiro2017budget}
Santiago Balseiro, Anthony Kim, Mohammad Mahdian, and Vahab Mirrokni.
\newblock Budget management strategies in repeated auctions.
\newblock In \emph{Proceedings of the 26th International World Wide Web
  Conference, Perth, Australia}, 2017.

\bibitem[Balseiro et~al.(2020{\natexlab{a}})Balseiro, Kim, Mahdian, and
  Mirrokni]{balseiro2020budget}
Santiago Balseiro, Anthony Kim, Mohammad Mahdian, and Vahab Mirrokni.
\newblock Budget-constrained incentive compatibility for stationary mechanisms.
\newblock In \emph{Proceedings of the 21st ACM Conference on Economics and
  Computation}, pages 607--608, 2020{\natexlab{a}}.

\bibitem[Balseiro et~al.(2020{\natexlab{b}})Balseiro, Lu, and
  Mirrokni]{balseiro2020dual}
Santiago Balseiro, Haihao Lu, and Vahab Mirrokni.
\newblock Dual mirror descent for online allocation problems.
\newblock In \emph{International Conference on Machine Learning}, pages
  613--628. PMLR, 2020{\natexlab{b}}.

\bibitem[Balseiro and Gur(2017)]{Balseiro:2017}
Santiago~R. Balseiro and Yonatan Gur.
\newblock Learning in repeated auctions with budgets: Regret minimization and
  equilibrium.
\newblock In \emph{Proceedings of the 2017 ACM Conference on Economics and
  Computation}, EC '17, pages 609--609, New York, NY, USA, 2017. ACM.
\newblock ISBN 978-1-4503-4527-9.
\newblock \doi{10.1145/3033274.3084088}.
\newblock URL \url{http://doi.acm.org/10.1145/3033274.3084088}.

\bibitem[Balseiro and Gur(2019)]{balseiro2019learning}
Santiago~R Balseiro and Yonatan Gur.
\newblock Learning in repeated auctions with budgets: Regret minimization and
  equilibrium.
\newblock \emph{Management Science}, 65\penalty0 (9):\penalty0 3952--3968,
  2019.

\bibitem[Borgs et~al.(2007)Borgs, Chayes, Immorlica, Jain, Etesami, and
  Mahdian]{borgs2007dynamics}
Christian Borgs, Jennifer Chayes, Nicole Immorlica, Kamal Jain, Omid Etesami,
  and Mohammad Mahdian.
\newblock Dynamics of bid optimization in online advertisement auctions.
\newblock In \emph{Proceedings of the 16th international conference on World
  Wide Web}, 2007.

\bibitem[Cary et~al.(2007)Cary, Das, Edelman, Giotis, Heimerl, Karlin, Mathieu,
  and Schwarz]{cray2007}
Matthew Cary, Aparna Das, Benjamin Edelman, Ioannis Giotis, Kurtis Heimerl,
  Anna~R. Karlin, Claire Mathieu, and Michael Schwarz.
\newblock Greedy bidding strategies for keyword auctions.
\newblock In Jeffrey~K. MacKie-Mason, David~C. Parkes, and Paul Resnick,
  editors, \emph{EC}, pages 262--271. ACM, 2007.
\newblock ISBN 978-1-59593-653-0.
\newblock URL
  \url{http://dblp.uni-trier.de/db/conf/sigecom/sigecom2007.html#CaryDEGHKMS07}.

\bibitem[Chen et~al.(2021)Chen, Kroer, and Kumar]{chen2021complexity}
Xi~Chen, Christian Kroer, and Rachitesh Kumar.
\newblock The complexity of pacing for second-price auctions, 2021.

\bibitem[Conitzer et~al.(2018)Conitzer, Kroer, Sodomka, and
  Stier-Moses]{conitzer2018pacing}
Vincent Conitzer, Christian Kroer, Eric Sodomka, and Nicolas~E. Stier-Moses.
\newblock Multiplicative pacing equilibria in auction markets.
\newblock In \emph{Conference on Web and Internet Economics (WINE'18)}, Oxford,
  UK, 2018.

\bibitem[Conitzer et~al.(2019)Conitzer, Kroer, Panigrahi, Schrijvers, Sodomka,
  Stier-Moses, and Wilkens]{ConitzerKPSSSW2019}
Vincent Conitzer, Christian Kroer, Debmalya Panigrahi, Okke Schrijvers, Eric
  Sodomka, Nicolas~E. Stier-Moses, and Chris Wilkens.
\newblock Pacing equilibrium in first-price auction markets.
\newblock In \emph{Proceedings of the 2019 ACM Conference on Economics and
  Computation}, EC '19, pages 587--587, New York, NY, USA, 2019. ACM.
\newblock ISBN 978-1-4503-6792-9.
\newblock \doi{10.1145/3328526.3329600}.
\newblock URL \url{http://doi.acm.org/10.1145/3328526.3329600}.

\bibitem[Davis et~al.(2011)Davis, Lii, and Politis]{davis2011remarks}
Richard~A Davis, Keh-Shin Lii, and Dimitris~N Politis.
\newblock Remarks on some nonparametric estimates of a density function.
\newblock In \emph{Selected Works of Murray Rosenblatt}, pages 95--100.
  Springer, 2011.

\bibitem[Dvoretzky et~al.(1956)Dvoretzky, Kiefer, and
  Wolfowitz]{dvoretzky1956asymptotic}
Aryeh Dvoretzky, Jack Kiefer, and Jacob Wolfowitz.
\newblock Asymptotic minimax character of the sample distribution function and
  of the classical multinomial estimator.
\newblock \emph{The Annals of Mathematical Statistics}, pages 642--669, 1956.

\bibitem[Feldman et~al.(2007)Feldman, Muthukrishnan, Pal, and
  Stein]{feldman2007budget}
Jon Feldman, S~Muthukrishnan, Martin Pal, and Cliff Stein.
\newblock Budget optimization in search-based advertising auctions.
\newblock In \emph{Proceedings of the 8th ACM conference on Electronic
  commerce}, 2007.

\bibitem[Flajolet and Jaillet(2017)]{flajolet2017real}
Arthur Flajolet and Patrick Jaillet.
\newblock Real-time bidding with side information.
\newblock In \emph{Proceedings of the 31st International Conference on Neural
  Information Processing Systems}, pages 5168--5178. Curran Associates Inc.,
  2017.

\bibitem[Gao et~al.(2021)Gao, Kroer, and Peysakhovich]{gao2021online}
Yuan Gao, Christian Kroer, and Alex Peysakhovich.
\newblock Online market equilibrium with application to fair division, 2021.

\bibitem[Goel et~al.(2010)Goel, Mahdian, Nazerzadeh, and Saberi]{goel2010gsp}
Ashish Goel, Mohammad Mahdian, Hamid Nazerzadeh, and Amin Saberi.
\newblock Advertisement allocation for generalized second-pricing schemes.
\newblock \emph{Oper. Res. Lett.}, 38\penalty0 (6):\penalty0 571--576, November
  2010.
\newblock ISSN 0167-6377.
\newblock \doi{10.1016/j.orl.2010.09.002}.
\newblock URL \url{http://dx.doi.org/10.1016/j.orl.2010.09.002}.

\bibitem[Hosanagar and Cherepanov(2008)]{hosanagar2008}
Kartik Hosanagar and Vadim Cherepanov.
\newblock Optimal bidding in stochastic budget constrained slot auctions.
\newblock In \emph{Proceedings 9th {ACM} Conference on Electronic Commerce
  (EC-2008), Chicago, IL, USA, June 8-12, 2008}, page~20, 2008.
\newblock \doi{10.1145/1386790.1386794}.
\newblock URL \url{http://doi.acm.org/10.1145/1386790.1386794}.

\bibitem[Jiang(2017)]{jiang2017uniform}
Heinrich Jiang.
\newblock Uniform convergence rates for kernel density estimation.
\newblock In \emph{International Conference on Machine Learning}, pages
  1694--1703. PMLR, 2017.

\bibitem[Kakade et~al.(2009)Kakade, Kalai, and Ligett]{kakade2009playing}
Sham~M Kakade, Adam~Tauman Kalai, and Katrina Ligett.
\newblock Playing games with approximation algorithms.
\newblock \emph{SIAM Journal on Computing}, 39\penalty0 (3):\penalty0
  1088--1106, 2009.

\bibitem[Karande et~al.(2013)Karande, Mehta, and Srikant]{karande2013pacing}
Chinmay Karande, Aranyak Mehta, and Ramakrishnan Srikant.
\newblock Optimizing budget constrained spend in search advertising.
\newblock In \emph{Proceedings of the Sixth ACM International Conference on Web
  Search and Data Mining}, WSDM '13, pages 697--706, New York, NY, USA, 2013.
  ACM.
\newblock ISBN 978-1-4503-1869-3.
\newblock \doi{10.1145/2433396.2433483}.
\newblock URL \url{http://doi.acm.org/10.1145/2433396.2433483}.

\bibitem[Lee et~al.(2013)Lee, Jalali, and Dasdan]{lee2013real}
Kuang-Chih Lee, Ali Jalali, and Ali Dasdan.
\newblock Real time bid optimization with smooth budget delivery in online
  advertising.
\newblock In \emph{Proceedings of the Seventh International Workshop on Data
  Mining for Online Advertising}, pages 1--9, 2013.

\bibitem[Liu and Hill(2020)]{liu2020moment}
Jia Liu and Shawndra Hill.
\newblock Moment marketing: Measuring dynamics in cross-channel ad
  effectiveness.
\newblock \emph{Available at SSRN 3670024}, 2020.

\bibitem[Ma et~al.(2019)Ma, Zhang, Xu, Liu, Chen, Xiao, Wang, and
  Wu]{ma2019large}
Xiaoyang Ma, Lan Zhang, Lan Xu, Zhicheng Liu, Ge~Chen, Zhili Xiao, Yang Wang,
  and Zhengtao Wu.
\newblock Large-scale user visits understanding and forecasting with deep
  spatial-temporal tensor factorization framework.
\newblock In \emph{Proceedings of the 25th ACM SIGKDD International Conference
  on Knowledge Discovery \& Data Mining}, pages 2403--2411, 2019.

\bibitem[Massart(1990)]{massart1990tight}
Pascal Massart.
\newblock The tight constant in the dvoretzky-kiefer-wolfowitz inequality.
\newblock \emph{The annals of Probability}, pages 1269--1283, 1990.

\bibitem[Mehta et~al.(2007)Mehta, Saberi, Vazirani, and
  Vazirani]{mehta2007adwords}
Aranyak Mehta, Amin Saberi, Umesh Vazirani, and Vijay Vazirani.
\newblock Adwords and generalized online matching.
\newblock \emph{Journal of the ACM (JACM)}, 54\penalty0 (5), 2007.

\bibitem[Nuara et~al.(2020)Nuara, Trovò, Gatti, and Restelli]{aless2020online}
Alessandro Nuara, Francesco Trovò, Nicola Gatti, and Marcello Restelli.
\newblock Online joint bid/daily budget optimization of internet advertising
  campaigns, 2020.

\bibitem[Ostrovsky and Schwarz(2011)]{ostrovsky2011reserve}
Michael Ostrovsky and Michael Schwarz.
\newblock Reserve prices in internet advertising auctions: A field experiment.
\newblock In \emph{Proceedings of the 12th ACM conference on Electronic
  commerce}, pages 59--60, 2011.

\bibitem[Parzen(1962)]{parzen1962estimation}
Emanuel Parzen.
\newblock On estimation of a probability density function and mode.
\newblock \emph{The annals of mathematical statistics}, 33\penalty0
  (3):\penalty0 1065--1076, 1962.

\bibitem[Rusmevichientong and Williamson(2006)]{rusmevichientong2006}
Paat Rusmevichientong and David~P. Williamson.
\newblock An adaptive algorithm for selecting profitable keywords for
  search-based advertising services.
\newblock In \emph{Proceedings 7th {ACM} Conference on Electronic Commerce
  (EC-2006), Ann Arbor, Michigan, USA, June 11-15, 2006}, pages 260--269, 2006.
\newblock \doi{10.1145/1134707.1134736}.
\newblock URL \url{http://doi.acm.org/10.1145/1134707.1134736}.

\bibitem[Statista(2020)]{statista2020}
Statista.
\newblock Online advertising revenue in the united states from 2000 to 2020,
  2020.
\newblock
  \url{https://www.statista.com/statistics/183816/us-online-advertising-revenue-since-2000/}.

\bibitem[Tran-Thanh et~al.(2012)Tran-Thanh, Chapman, Rogers, and
  Jennings]{tran2012knapsack}
Long Tran-Thanh, Archie Chapman, Alex Rogers, and Nicholas~R Jennings.
\newblock Knapsack based optimal policies for budget--limited multi--armed
  bandits.
\newblock In \emph{Twenty-Sixth AAAI Conference on Artificial Intelligence},
  2012.

\end{thebibliography}
